\documentclass[11pt]{article}

       \newcommand{\Mc}{ {\mathcal{M}} }

       \newcommand{\Pc}{ {\mathcal{P}} }
       
       \newcommand{\Rc}{ {\mathcal{R}} }
       
       \newcommand{\Tc}{ {\mathcal{T}} }

  \newcommand{\nn}{\nonumber}

\usepackage{graphicx}
\usepackage{epstopdf, epsfig}
\usepackage{amsfonts}
\usepackage{amssymb}
\usepackage[T1]{fontenc}
\usepackage{mathtools}
\usepackage{amsfonts}
\usepackage{dcolumn}
\usepackage{bm}
\usepackage[export]{adjustbox}
\usepackage{wrapfig}
\usepackage{lipsum}
\usepackage{authblk}
\usepackage[font=small,labelfont=bf]{caption}
\usepackage{booktabs}
\allowdisplaybreaks
\usepackage[titletoc,toc,title]{appendix}	

\usepackage[style=ieee, citestyle=numeric-comp, backend=biber]{biblatex}

\usepackage{blindtext}
\usepackage[a4paper,
            bindingoffset=0.2in,
            left=1in,
            right=1in,
            top=1in,
            bottom=1in,
            footskip=.25in]{geometry}

\usepackage{hyperref}
\hypersetup{
    colorlinks=true,
    linkcolor=red,
    filecolor=magenta,      
    urlcolor=cyan,
    citecolor = cyan,
    }
 
\title{Multi-soliton solutions and data-driven discovery of higher-order Burgers' hierarchy equations with physics informed neural networks} 
\author[1,2]{D. A. Kaltsas\thanks{kaltsas.d.a@gmail.com}}
\author[2]{L. Magafas} 
\author[3]{P. Papadopoulou} 
\author[1]{G. N. Throumoulopoulos} 
\affil[1]{Department of Physics, University of Ioannina, Ioannina GR 451 10, Greece}
\affil[2]{Department of Informatics, Democritus University of Thrace, Kavala GR 654 04, Greece}
\affil[3]{Department of Physics, Democritus University of Thrace, Kavala GR 654 04, Greece}

\date{}

\begin{document}
\maketitle
 \vspace{-10mm}
\begin{abstract}
The Burgers hierarchy consists of nonlinear evolutionary partial differential equations (PDEs) with progressively higher-order dispersive and nonlinear terms. Notable members of this hierarchy are the Burgers equation and the Sharma-Tasso-Olver equation, which are widely applied in fields such as plasma physics, fluid mechanics, optics, and biophysics to describe nonlinear waves in inhomogeneous media. Various soliton and multi-soliton solutions to these equations have been identified and the fission and fusion of solitons have been studied using analytical and numerical techniques. Recently, deep learning methods, particularly Physics-Informed Neural Networks (PINNs), have emerged as a new approach for solving PDEs. These methods use deep neural networks to minimize PDE residuals while fitting relevant data. Although PINNs have been applied to equations like Burgers' and Korteweg-de Vries, higher-order members of the Burgers hierarchy remain unexplored in this context. In this study, we employ a PINN algorithm to approximate multi-soliton solutions of linear combinations of equations within the Burgers hierarchy. This semi-supervised approach encodes the PDE and relevant data, determining PDE parameters and resolving the linear combination to discover the PDE that describes the data. Additionally, we employ gradient-enhanced PINNs (gPINNs) and a conservation law, specific to the generic Burgers' hierarchy, to improve training accuracy. The results demonstrate the effectiveness of PINNs in describing multi-soliton solutions within the generic Burgers' hierarchy, their robustness to increased levels of data noise, and their limited yet measurable predictive capabilities. They also verify the potential for training refinement and accuracy improvement using enhanced approaches in certain cases, while enabling the discovery of the PDE model that describes the observed solitary structures.

\end{abstract}


\section{Introduction}
\label{Sec_I}
Solitons are dispersionless solutions to evolutionary nonlinear Partial Differential Equations (PDEs) that travel long distances at a constant speed without changing shape. These soliton equations have extensive applications in plasma physics, fluid dynamics, astrophysics, statistical physics, optics, and biophysics. Solitons have been experimentally observed in shallow waters, space plasmas, Bose-Einstein condensates, optical fibers, and DNA lattices (e.g., see \cite{Arora2022}). Finding new analytical and numerical solutions, conservation laws, and symmetries of soliton equations is crucial for understanding soliton behavior in various physical systems.

Typical examples of nonlinear evolutionary soliton equations belong to a general family of integrable PDEs called Burgers' hierarchy \cite{Gerdjikov2008}. In $(1+1)$ dimensions this hierarchy is given by:
\begin{eqnarray}
    \partial_t u +\alpha \partial_x \left(\partial_x + u\right)^nu=0\,,\quad n=0,1,2,...\,,
    \label{Burgers'_hierarchy_1}
\end{eqnarray}
where $u=u(x,t)$ and $\alpha$ is a constant parameter. The first two members of the Burgers hierarchy are the linear advection and the Burgers \cite{Burgers1948} equations, obtained for $n=0$ and $n=1$, respectively:
\begin{eqnarray}
    \partial_t u +\alpha \partial_x u =0\,, \label{advection}\\ 
    \partial_t u +\alpha(2u \partial_x u + \partial_{xx}u)=0\,. \label{Burgers'}
\end{eqnarray}
Next in this hierarchy, for $n=2$, is the Sharma-Tasso-Olver (STO) equation:
\begin{eqnarray}
    \partial_t u +\alpha\left[3u\partial_{xx}u+3u^2\partial_xu+3(\partial_x u)^2+\partial_{xxx}u\right]=0\,, \label{STO}
\end{eqnarray}
derived as the first odd member of Burgers' hierarchy by H. Tasso in \cite{Tasso1976} and \cite{Sharma1977} and appeared also in \cite{Olver1977}. Considering $n=3$, the 4th order Burgers' equation emerges:
\begin{eqnarray}
    \partial_t u + \alpha\left(10 \partial_x u\partial_{xx}u+4u\partial_{xxx}u+12 u (\partial_xu)^2+6u^2\partial_{xx}u+4u^3\partial_xu+\partial_{xxxx}u\right)=0\,.\label{4th_order_Burgers'}
\end{eqnarray}
Here we will consider linear combinations of equations \eqref{Burgers'}--\eqref{4th_order_Burgers'} to describe solitary wave interactions such as soliton fission and fusion of triple solitons. To this end we consider a generic Burgers' hierarchy of the form:
\begin{eqnarray}
    \partial_t u +\sum_{n=0}^N\alpha_n \partial_x \left(\partial_x + u\right)^nu=0\,, \quad N=0,1,2,...\,. \label{generic_Burgers'_hierarchy}
\end{eqnarray}
Setting $\alpha_0=0$ and $N=2$, results in a combination of the STO and Burgers' equation  known as the Sharma-Tasso-Olver-Burgers' (STOB) equation (see for example \cite{Hosseini2022, Yan2020}). For $N=3$ we obtain a combination of the STOB equation with the fourth-order Burgers' equation. This is a fourth-order nonlinear PDE that includes the nonlinear and dispersive terms of the Burgers equation, the STO equation, and the fourth-order Burgers' equation \eqref{4th_order_Burgers'}. Evidently, for $\alpha_2=\alpha_3=0$ this simplifies to the Burgers equation \eqref{Burgers'}. When $\alpha_1=\alpha_3=0$ it reduces to the STO equation \eqref{STO} and if $\alpha_1=\alpha_2=0$ it becomes the fourth-order Burgers' equation \eqref{4th_order_Burgers'}.

There have been numerous analytical solution methods for resolving nonlinear evolutionary PDEs. To this end one can use various transformations to obtain reductions of the PDEs such as the Miura \cite{Miura1968}, the Cole-Hopf \cite{Cole1951,Hopf1950}, the B\"acklund (e.g. \cite{Rogers1982}) and other types of transformations, similarity reductions obtained from Lie-point and generalized symmetries, the Hirota bilinear \cite{Hirota1971}, the tanh \cite{Malfliet2004}, sine-cosine \cite{Wazwaz2004}, the Kudryashov \cite{Kudryashov2012}, the exponential function  \cite{He2006} and many other methods. Plenty of these methods have been applied for equations stemming from the Burgers hierarchy. For  example, Kudryashov and Sinelshchikov in \cite{Kudryashov2009} used the Cole-Hopf transformation to find solutions to equations from the Burgers hierarchy; Wazwaz in \cite{Wazwaz2007} used the tanh method for the STO equation; in \cite{Wazwaz2010} the Hirota bilinear method was used for the Burgers hierarchy and in \cite{Wazwaz2018} the Cole-Hopf transform combined with a simplified Hirota method to derive solutions to a two-mode STO and a two-mode fourth order Burgers' equation. Kudryashov and exponential methods have been applied recently in \cite{Hosseini2022} to the STOB equation.

Numerical methods are also employed, including the finite difference and finite element methods (e.g., \cite{Hassanien2005,Ozis2003}), spectral methods \cite{Boyd1989}, explicit and implicit multistep methods (e.g., \cite{Wang2008}), the lattice Boltzmann method \cite{Feng2023}, and others. In recent years, the booming interest in numerical methods based on physics-informed neural networks (PINNs) \cite{Raissi2019,Blechschmidt2021,Cuomo2022} and deep learning (DL) has led researchers to seek neural network solutions to various soliton equations, including the Burgers and Korteweg-de Vries (KdV) equations \cite{Li2020,Guo2020,Cui2022}. However, none of the existing studies have considered higher-order members of the Burgers hierarchy, such as the STO equation and the fourth-order Burgers' equation. 

{ DL methods involve training deep neural networks (DNNs) to approximate solutions to PDEs by optimizing their internal parameters. Although they are generally less efficient than traditional numerical methods in terms of computational time and accuracy for solving direct problems, they offer significant advantages in specific contexts. Traditional methods rely on explicit formulas or matrix inversions and generally solve PDEs more quickly. In contrast, the optimization process for training DNNs is typically slower but excels in tasks like data assimilation or multi-query problems. DL methods can seamlessly integrate sparse experimental or observational data with model equations, enabling efficient discovery of parametric values and facilitating the identification of mathematical models that describe observed phenomena. This approach also helps to uncover physical mechanisms, such as dispersion or diffusion, that may not be immediately evident from observations.

Moreover, the optimization-based training of DNNs allows for the convenient imposition of constraints, e.g. conservation laws, by introducing appropriate penalty terms to the target function. This approach is exploited in the present work. While traditional numerical methods can also respect conservation laws--such as those that exploit the multisymplectic \cite{Bridges2001} structure of certain PDEs (e.g., the KdV equation, as described in \cite{Ascher2004})--their implementation often requires deriving the specific multi-symplectic structure, which can be complex and non-trivial.  

DL methods also demonstrate advantages in multi-query problems, where the same equation must be solved for various parameter values. In these scenarios, DNNs can learn the parametric dependence of solutions by training on instances of the PDE with different parameter sets. Once trained, the model can produce solutions for new parameter values at inference time, making the DL approach more efficient than traditional numerical methods for such applications. Exploring the application of DL methods to the Burgers' hierarchy in multi-query scenarios will be the topic of future research.}

In this work we consider a combination of the members $n=1,2,3$ of the Burgers hierarchy and use semi-supervised learning to resolve this combination obtained from \eqref{generic_Burgers'_hierarchy} with $N=3$. We use data obtained by perturbing  exact solutions of \eqref{generic_Burgers'_hierarchy} evaluated at randomly sampled points of the spatio-temporal domain, with gaussian noise. Eq. \eqref{generic_Burgers'_hierarchy} with unknown parameters $\alpha_n$, $n=1,2,3$, is used to train feed-forward, fully connected DNNs to fit the data and satisfy the equation simultaneously, thus inferring the specific values of the parameters $\alpha_n$.  Although handling such inverse problems with PINNs and DL methods has been extensively studied, this is the first time the technique is utilized for resolving combinations of equations in the Burgers hierarchy. This suggests that this technique can be used for the resolution of such hierarchies, identifying the suitable PDE models that can describe observed soliton structures in nature. In this respect, this work resembles more closely the very recent research on data-driven soliton solutions in \cite{Fang2022,Wang2024a}, which concern the KdV and modified KdV equations, and \cite{Wang2023,Wang2024b}, which focus on the Kaup–Kuperschmidt equation and multi-component and high-dimensional coupled nonlinear partial differential equations, respectively.

An additional feature in this study is the combined implementation of gradient-enhanced PINNs (gPINNs), introduced in \cite{Yu2022}, with conservation-law constrained PINNs  (e.g., see \cite{Fang2022}). In the gPINN algorithm, gradient information of the PDE residual is utilized by embedding the gradient of the PDE with respect to its independent variables into the loss function which is minimized for the network training. In some cases, this improves the training accuracy, provided that the weight of the additional gradient term in the loss function is carefully selected. Additionally, we impose a conservation law admitted by the generic Burgers' hierarchy \eqref{generic_Burgers'_hierarchy} to improve training accuracy. { This approach has been employed in previous works, such as \cite{Fang2022} and similar studies \cite{Wu2022, Jagtap2020, Gurieva2022}, using neural networks termed Conservation-Law Constrained Neural Networks (CLCNNs) or Conservative Physics-Informed Neural Networks (cPINNs). Notably, \cite{Fang2022} and \cite{Wu2022} focus on soliton solutions of nonlinear PDEs. These studies found that imposing PDE conservation laws enhances solution and parameter inference accuracy, particularly when training data are limited \cite{Fang2022}. However, this improvement comes at the expense of increased computational costs and longer training times \cite{Wu2022}.

In the present work, we corroborate these findings, showing that conservation-law constrained PINNs (termed here as cl-PINNs) can also be advantageous for the model studied here, particularly when training data is limited, though at the cost of longer training times. Additionally, we compare this enhanced method with the gPINN method and a novel combination of the two (termed cl-gPINN), which has not been reported in the literature. These comparisons indicate that the gPINN algorithm is more efficient, providing comparable or greater accuracy improvements than the other two enhanced methods with minimal increases in training time.}

The rest of the paper is organized as follows: In Section \ref{sec_II}, we present the Cole-Hopf transform, which linearizes the Burgers hierarchy and allows for the straightforward construction of exact solitary solutions. Using this transform, we derive new exact solution classes for a linear combination of the lower-order members of this hierarchy, including derivatives up to the fourth order. In Section \ref{sec_III}, we introduce PINNs and enhanced PINNs, along with the associated algorithms for resolving the generic Burgers' hierarchy \eqref{generic_Burgers'_hierarchy}. In Section \ref{sec_IV}, we discuss the numerical experiments featuring the fission and fusion of solitary waves, we perform a convergence and noise analysis of the baseline PINN algorithm and present cases where the enhanced algorithms provide improved accuracy. Finally, in Section \ref{sec_V}, we summarize the results and conclusions of the study.


\section{The Cole-Hopf transform and multi-soliton solutions}
\label{sec_II}
The generic Burgers' hierarchy \eqref{generic_Burgers'_hierarchy} can be linearized by the Cole-Hopf transform \cite{Hopf1950,Cole1951}
\begin{eqnarray}
u = \frac{\partial_x \phi}{\phi}\,, \label{cole_hopf}
\end{eqnarray}
where $\phi(x,t)$ is an auxiliarry function. Substituting \eqref{cole_hopf} into \eqref{generic_Burgers'_hierarchy} we get:
\begin{eqnarray}
    \frac{\partial_{xt}\phi}{\phi}-\frac{\partial_x\phi \partial_t\phi}{\phi^2}+\sum_{n=0}^N \alpha_n\left(\frac{\partial_x^{(n+2)}\phi}{\phi}-\frac{{\partial_x\phi}\partial_x^{(n+1)}\phi
    }{\phi^2}\right)=0 \,. \label{bilinear}
\end{eqnarray}
Equation \eqref{bilinear} can be integrated with respect to $x$:
\begin{eqnarray}
\partial_t \phi+ \sum_{n=0}^N \alpha_n \partial_{x}^{(n+1)}\phi=\beta(t)\phi\,, \quad N=1,2,...\,,\label{linearized_Burgers'_hierarchy}
\end{eqnarray} 
which is the linearized version of generic Burgers' hierarchy \eqref{generic_Burgers'_hierarchy}. Setting $\beta(t)=0$ allows considering solutions of the form $\phi = e^{\omega t+ k x}$, thus obtaining the following dispersion relation:
\begin{eqnarray}
    \omega  = - \sum_{n=0}^N \alpha_n k^{n+1}\,. \label{disp_rel}
\end{eqnarray}
A general multi-soliton solution to \eqref{linearized_Burgers'_hierarchy} can be written as 
\begin{eqnarray}
    \phi(x,t) = \phi_0 + \sum_{i=1}^K c_i e^{\omega_i t+k_i x+\xi_i}\,, \label{gen_soliton_phi_1}
\end{eqnarray}
where $\xi_i$ are constant phases. Therefore, general K-soliton solutions to the generalized Burgers' equation of the form \eqref{generic_Burgers'_hierarchy} for some fixed $N$ are given by:
\begin{eqnarray}
    u(x,t) = \frac{\sum_{i=1}^K c_i k_i e^{\omega_i t+k_i x+\xi_i}}{\phi_0 + \sum_{i=1}^K c_i e^{\omega_i t+k_i x+\xi_i}}\,. \label{gen_soliton_1}
\end{eqnarray}
Additional classes of solutions can be obtained by the bilinear approach used in \cite{Hossen2021} for the Burgers equation. This approach essentially consists of solving \eqref{bilinear} before integrating over $x$. Thus, we expect that these additional classes of solutions can be obtained from the linearized equation \eqref{linearized_Burgers'_hierarchy} for $\beta\neq 0$. Here we consider $\beta=const.\neq 0$ but this constant value should be related to the constant wavenumbers $k_1,...,k_K$, i.e. $\beta=\beta(\textbf{k})=const.$ where $\textbf{k}=(k_1,...,k_K)^T$. In this case we can seek classes of K-soliton solutions of the form:
\begin{eqnarray}
    \phi(x,t) = \sum_{j=1}^K \sum_{1\leq i_1<i_2\hdots <i_j\leq K}c_{i_1i_2\hdots i_j} \prod_{m=1}^{j}e^{\zeta_{i_m}}\,,
    \label{gen_soliton_phi_2}
\end{eqnarray}
where $\zeta_i = \omega_i t + k_i x$. For example, two- and three-soliton solutions are described by:
\begin{eqnarray}
    \phi(x,t) &=& c_1 e^{\zeta_1} + c_2 e^{\zeta_2} + c_{12} e^{\zeta_1+\zeta_2}\,, \label{phi_2_soliton_2}\\
    \phi(x,t) &=& c_1 e^{\zeta_1} + c_2 e^{\zeta_2}+ c_3 e^{\zeta_3} + c_{12} e^{\zeta_1+\zeta_2}+c_{13}e^{\zeta_1+\zeta_3}+c_{23}e^{\zeta_2+\zeta_3}+c_{123} e^{\zeta_1+\zeta_2+\zeta_3}\,,\label{phi_3_soliton_2}
\end{eqnarray}
respectively, generalizing the ansatzes considered in \cite{Hossen2021}. Substituting the general solution \eqref{gen_soliton_phi_2} into \eqref{linearized_Burgers'_hierarchy} we can determine $\beta(\textbf{k})$ along with the parameters $\omega_i$, $c_i$, $c_{i_1i_2}$,...,$c_{i_1\hdots i_K}$ by setting the coefficients of the different exponential functions equal to zero.

In this work we resolve Burgers' hierarchy up to 4th order, i.e. we consider Eq. \eqref{generic_Burgers'_hierarchy} with $N=3$ and $a_0=0$, approximating  three-soliton solutions with deep neural networks. Exact soliton solutions are perturbed and used as data for training the networks while simultaneously the PDE residual is minimized. This minimization is enhanced by embedding into the loss function the gradient of the PDE residual and a conservation law of the generic Burgers' hierarchy. The associated conserved quantity is given by:
\begin{eqnarray}
    C = \int_{-\infty}^{\infty}\phi(x,t)dx\,. \label{conserved_quantity}
\end{eqnarray}

It can be readily verified that $d\mathcal{C}/dt=0$ for specific $\beta(t)$ using \eqref{linearized_Burgers'_hierarchy}:
\begin{eqnarray}
    \frac{d C}{dt} = \frac{d}{dt}\int_{-\infty}^{+\infty} \phi(x,t) dx = \int_{-\infty}^{+\infty} \partial_t\phi(x,t) dx \nn\\= \beta(t)\int_{-\infty}^{+\infty} \phi(x,t)dx- \sum_{n=0}^{N}\alpha_n \int_{-\infty}^{+\infty} \partial_x^{(n+1)}\phi(x,t)dx \nn \\ 
    =\beta(t)\int_{-\infty}^{+\infty}\phi(x,t)dx-\sum_{n=0}^N\alpha_n \big[\partial_x^{(n)}\phi(x,t)\big]_{-\infty}^{+\infty}\,.
\end{eqnarray}
Thus, $C$ is a constant of motion if and only if the arbitrary function $\beta(t)$ cancels the boundary term:
$$\beta(t)\int_{-\infty}^{+\infty}\phi(x,t)dx=\sum_{n=0}^N\alpha_n \big[\partial_x^{(n)}\phi(x,t)\big]_{-\infty}^{+\infty}\,.$$
Hence, if the boundary term vanishes (e.g., \(\phi\) and its derivatives decay at infinity), then \(\beta(t) = 0\) is necessary for conservation.

If, however, 
$$\sum_{n=0}^N\alpha_n \big[\partial_x^{(n)}\phi(x,t)\big]_{-\infty}^{+\infty}=0\,,$$ but $\beta(t)\neq 0$, then: 
\begin{eqnarray}
    C(t) = \int_{-\infty}^{+\infty}\phi(x,t)dx=C_0 \, exp{\left(\int_{-\infty}^t \beta(s)ds\right)} \,,
\end{eqnarray}
where $C_0=\int_{-\infty}^{+\infty}\phi(x,-\infty)dx$. In this case, the conserved quantity is:
\begin{eqnarray}
    C_0 =  exp{\left(-\int_{-\infty}^t \beta(s)ds\right)}\int_{-\infty}^{+\infty}\phi(x,t)dx \,.
\end{eqnarray}
The original $C$ is a special case of $C_0$ which corresponds to $\beta(t)=0$. In what follows we consider the case $\beta(t)=0$, therefore $C_0=C$.

Now, from the Cole-Hopf transform \eqref{cole_hopf} one has:
\begin{eqnarray}
    \phi(x,t) = \phi_0\exp{\int_{-\infty}^{x}u(x',t,)dx'}\,,
\end{eqnarray}
where $\phi_0=\phi(-\infty,t)$ assumed to be constant. Thus, the conserved quantity $C$ is written in terms of $u(x,t)$ as follows:
\begin{eqnarray}
    C=\phi_0\int_{-\infty}^{+\infty} \exp{\int_{-\infty}^{x}u(x',t,)dx'} dx\,. \label{CL_u}
\end{eqnarray}

\subsection{Double soliton solutions}
To derive a first class of double soliton solutions {to Eq. \eqref{generic_Burgers'_hierarchy} with $N=3$}, we assume $\beta=0$ and $\phi_0=1$. In this case Eq. \eqref{gen_soliton_1} with $K=2$ describes a class of solutions with dispersion relation $\omega_i = -\alpha_1 k_i^2  -\alpha_2 k_i^3 - \alpha_3 k_i^4$\,, $i=1,2$, i.e.:
\begin{eqnarray}
    u_{2}^{(1)}(x,t)=\frac{c_1k_1 e^{\omega_1 t + k_1  x+\xi_1}+c_2 k_2 e^{\omega_2 t + k_2 x+\xi_2}}{1+c_1e^{\omega_1 t + k_1  x+\xi_1}+c_2e^{\omega_2 t + k_2 x+\xi_2}}\,,\nn \\
    \omega_i = -\alpha_1 k_i^2 -\alpha_2 k_i^3-\alpha_3 k_i^{4}\,,\quad i=1,2\,,  \label{double_soliton_1}
\end{eqnarray}
where the subscript $2$ indicates the number of solitons and the superscript $(1)$ indicates the class. Two alternative classes can be found for $\phi_0 =0$ and $\beta\neq 0$. These classes are obtained from an ansatz  for the auxiliary   function  $\phi$  of the form \eqref{gen_soliton_phi_2},  resulting  in:
\begin{eqnarray}
    u_{2}^{(2)}(x,t) = \frac{c_1 k_1 e^{\omega_1 t+k_1 x+\xi_1}+c_2 k_2 e^{\omega_2 t +k_2 x+\xi_2}}{c_1  e^{\omega_1 t+k_1 x+\xi_1}+c_2 e^{\omega_2 t +k_2 x+\xi_2}}\,, \nn \\ 
    \omega_2= \alpha_1 k_1^2 + \alpha_2 k_1^3 +\alpha_3 k_1^4-\alpha_1 k_2^2 - \alpha_2 k_2^3 -\alpha_3 k_2^4+\omega_1\,, \label{double_soliton_2}
\end{eqnarray}
and
\begin{eqnarray}
    u_{2}^{(3)}(x,t) =  \frac{c_1 k_1 e^{\omega_1 t+k_1 x+\xi_1}+c_2 k_2 e^{\omega_2 t +k_2 x+\xi_2}+c_{12}(k_1+k_2)exp^{(\omega_1+\omega_2)t+(k_1+k_2)x+\xi_1+\xi_2}}{c_1  e^{\omega_1 t+k_1 x+\xi_1}+c_2  e^{\omega_2 t +k_2 x+\xi_2}+c_{12}exp^{(\omega_1+\omega_2)t+(k_1+k_2)x+\xi_1+\xi_2}}\,, \nn \\ 
    \omega_1 = -k_1 [\alpha_1 (k_1 + 2 k_2) + \alpha_3 (k_1 + 2 k_2) (k_1^2 + 2 k_1 k_2 + 2 k_2^2) + 
   \alpha_2 (k_1^2 + 3 k_1 k_2 + 3 k_2^2)]\,,\nn\\
   \omega_2 = -k_2 [\alpha_1 (2 k_1 + k_2) + \alpha_3 (2 k_1 + k_2) (2 k_1^2 + 2 k_1 k_2 + k_2^2) + 
   \alpha_2 (3 k_1^2 + 3 k_1 k_2 + k_2^2)]\,. \label{double_soliton_3}
\end{eqnarray}

\subsection{Triple soliton solutions}
A first class of triple solitons can be  found considering $\beta=0$. In this case, from Eq. \eqref{gen_soliton_1} with $K=3$ and $\phi_0=1$ and the dispersion relation \eqref{disp_rel}, we obtain the following solution:
\begin{eqnarray}
    u_{3}^{(1)}(x,t) = \frac{c_1k_1 e^{\omega_1 t  +k_1x +\xi_1 }+c_2k_2e^{\omega_2  t + k_2 x +\xi_2}+c_3k_3e^{\omega_3 t + k_3 x +\xi_3}}{1+ c_1e^{\omega_1 t  +k_1x +\xi_1 }+c_2e^{\omega_2 t  +k_2x +\xi_2 }+c_3e^{\omega_3 t  +k_3x +\xi_3 }}\,,   \nn \\ 
    \omega_i = -\alpha_1 k_i^2 - \alpha_2 k_i^3 -\alpha_3 k_i^4\,, \quad i=1,2,3\,. \label{triple_soliton_1}
\end{eqnarray}
Alternative classes are obtained for $\phi_0=0$ using the ansatz \eqref{gen_soliton_phi_2} with $K=3$ resulting in solutions of the form:
\begin{eqnarray}
   && u_{3}^{(m)} = \frac{c_1k_1 e^{\zeta_1} + c_2 k_2 e^{\zeta_2} + c_3k_3 e^{\zeta_3}+ c_{12}(k_1+k_2)e^{\zeta_1+\zeta_2}+c_{13} (k_1+k_3)e^{\zeta_1+\zeta_3}}{c_1e^{\zeta_1} + c_2 e^{\zeta_2} + c_3 e^{\zeta_3}+ c_{12}e^{\zeta_1+\zeta_2}+c_{13} e^{\zeta_1+\zeta_3}+c_{23} e^{\zeta_2+\zeta_3}+c_{123}e^{\zeta_1+\zeta_2+\zeta_3}}\nn\\
    && \hspace{-10mm}+\frac{c_{23}(k_2+k_3)   e^{\zeta_2+\zeta_3}+ c_{123}(k_1+k_2+k_3)e^{\zeta_1+\zeta_2+\zeta_3}}{c_1e^{\zeta_1} + c_2 e^{\zeta_2} + c_3 e^{\zeta_3}+ c_{12}e^{\zeta_1+\zeta_2}+c_{13} e^{\zeta_1+\zeta_3}+c_{23} e^{\zeta_2+\zeta_3}+c_{123}e^{\zeta_1+\zeta_2+\zeta_3}}\,, \; m=2,...,6\,.\label{triple_soliton_m}
\end{eqnarray}
The general class \eqref{triple_soliton_m} describes 5 sub-classes, which are documented in Appendix \ref{appendix_A}.  

For the semi-supervised training of the DNNs we will use the class $u_{3}^{(1)}$. The rest are reported not only for completeness but also because they have not been documented elsewhere in the literature for the fourth-order system \eqref{generic_Burgers'_hierarchy} with $N=3$.

\section{PINNs for the generic Burgers' hierarchy}
\label{sec_III}
The PINN algorithm for addressing boundary and initial value problems in physics has  been described in several works  over the last years, e.g. in \cite{Raissi2019,Blechschmidt2021,Cuomo2022}. Here we exploit this algorithm to solve  equations of the form \eqref{generic_Burgers'_hierarchy} using data to infer the parameter values $\alpha_n$. To this end we approximate the unknown function $u(x,t)$ using a fully connected, feed-forward DNN with $\ell$ hidden layers:
\begin{eqnarray}
    u_{net}(x,t;\boldsymbol{p}) = \boldsymbol{W}_\ell \circ \boldsymbol{f}  \circ \boldsymbol{W}_{\ell-1}\circ\hdots \circ\boldsymbol{W}_1\circ\boldsymbol{f}\circ\boldsymbol{W}_0\cdot \boldsymbol{z}\,, \label{neural_net}
\end{eqnarray}
where $\boldsymbol{z}=(x,t)^T$, $\boldsymbol{W}_i$, $i=0,...,\ell$ are affine transformations and $f$ is a nonlinearity. { Equation \eqref{neural_net}  shows that a feed-forward DNN  is essentially  an alternating sequence of a nonlinear function $f$ and an affine transformation  $W$}. In this study we select $f(x)=tanh(x)$ and $f(x)=1/(1+e^{-x})$. The vector $\boldsymbol{p}$ comprises the weight and bias parameters of the affine transformations  $\boldsymbol{W}_i$.

\subsection{The PINN algorithm}
The neural network \eqref{neural_net} is trained  to satisfy equation \eqref{generic_Burgers'_hierarchy} (PDE constraint) with $N=3$  and a set of ``experimental'' {noisy solution data (data constraint), that are available on a set of structured points $\mathcal{T}=\{\mathcal{T}_i=(x_i,t_i), i=1,...,M_\mathcal{T}\}$, within a domain $D = [x_{min},x_{max}]\times[t_{min},t_{max}]\subset \mathbb{R}^2$. The experimental data are generated by perturbing the solution \eqref{gen_soliton_1} with small noise as described subsequently in subsection \ref{subsec_3.3}  and quantified in the subsections \ref{subsec_4.1} and \ref{subsec_4.2}. To impose the PDE constraint, the PDE residual is evaluated at a set $\mathcal{P}=\{\mathcal{P}_i=(x_i,t_i), i=1,...,M_\mathcal{P}\}$ of random collocation points, the number of which is specified also in the subsequent numerical experiments. These points are distributed within the domain $D$}. The training process involves minimizing an appropriate target function $L(\boldsymbol{p},\boldsymbol{\alpha})$,  known as the loss function, with respect to the trainable parameters $(\boldsymbol{p},\boldsymbol{\alpha})$:
\begin{eqnarray}
\underset{(\boldsymbol{p},\boldsymbol{\alpha})}{\mathrm{argmin}}\,L(\boldsymbol{p},\boldsymbol{\alpha})\,,
\end{eqnarray}
where the loss function is defined as:
\begin{eqnarray}
    L = w_{e} L_{e} + w_{d} L_{d}\,, \label{loss_tot}
\end{eqnarray}
with
\begin{eqnarray}
    L_{e} = \frac{1}{M_\mathcal{P}} \sum_{\mathcal{P}_i\in \mathcal{P}}\bigg|\partial_t u_{net} +\sum_{n=0}^{N}\alpha_n \partial_x \left(\partial_x + u_{net}\right)^nu_{net}\bigg|^2\,, \label{loss_eq} \\ 
    L_{d} = \frac{1}{M_\mathcal{T}} \sum_{\mathcal{T}_i\in \mathcal{T}}\bigg|u_{net}(x_i,t_i;\boldsymbol{p})-u_{dat}(x_i,t_i)\bigg|^2\,. \label{loss_dat}
\end{eqnarray}
The partial derivatives in \eqref{loss_eq} are calculated using automatic differentiation and the backpropagation algorithm with PyTorch's autograd engine. The parameters $w_{e}$, $w_{d}$ in \eqref{loss_tot} are weights that quantify the significance of each term in the training process. Note that we do not introduce additional terms for the imposition of the initial and boundary  conditions, as these can be incorporated into the term $L_d$. 

\begin{figure}[h!]
    \centering
    \includegraphics[scale=0.5]{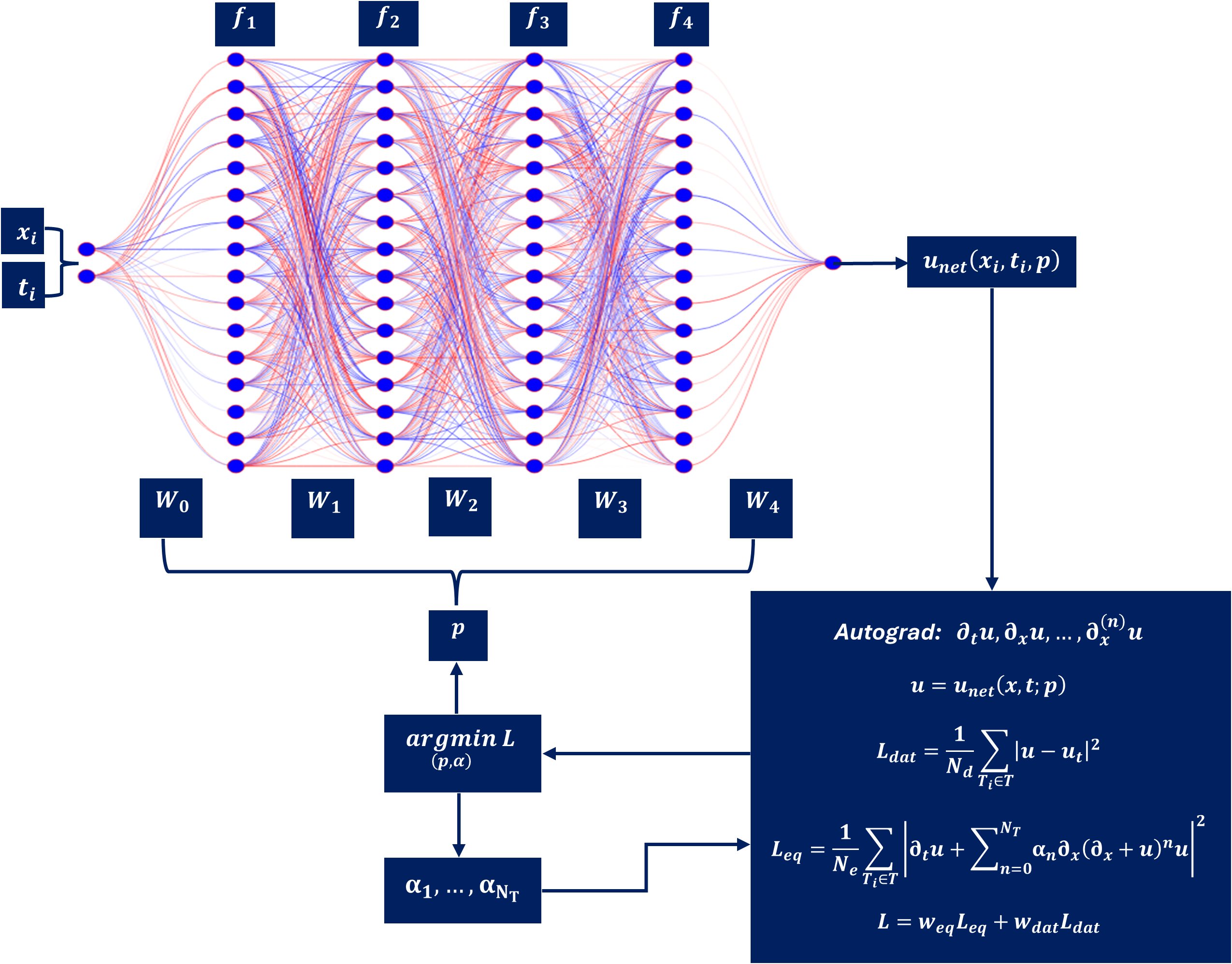}
    \caption{The PINN algorithm for the approximation of solutions and data-driven discovery of equations within the hierarchy \eqref{generic_Burgers'_hierarchy}.}
    \label{fig_train}
\end{figure}

\subsection{The gPINN algorithm}
In \cite{Yu2022}, Yu et al. proposed the gradient-enhanced PINNs (gPINNs), which improved training accuracy compared to standard PINNs in approximating the solutions of PDEs for the same number of training epochs. This improvement is achieved by incorporating the gradient of the PDE residual with respect to the independent variables into the loss function. However, this enhancement increases the overall computational cost and runtime, since gPINNs involve loss terms with higher-order derivatives. For optimized parameters, the trade-off between accuracy and computational cost can result in improved efficiency. In the gPINN approach an additional gradient loss term is included in the total loss. This term is:
\begin{eqnarray}
    L_g = \frac{1}{M_\mathcal{G}} \sum_{\mathcal{G}_i\in \mathcal{G}}\bigg|\frac{\partial \mathcal{R}(u_{net}(x_i,t_i))}{\partial x_i}  \bigg|^2 + \frac{1}{M_\mathcal{G}} \sum_{\mathcal{G}_i\in \mathcal{G}}\bigg|\frac{\partial \mathcal{R}(u_{net}(x_i,t_i))}{\partial t_i}  \bigg|^2 \,, \label{grad_constr}
\end{eqnarray}
where $\mathcal{R}$ is the PDE residual and $\mathcal{G} = \{\mathcal{G}_i=(x_i,t_i), i=1,...,M_{\mathcal{G}}\}$ is a set of points within the spatio-temporal computational domain $D$ used for the evaluation of $L_g$. In our case
\begin{eqnarray}
    \mathcal{R} = \partial_t u(x,t) {+} \sum_{n=0}^{N} \partial_x (\partial_x + u)^n u\,,
\end{eqnarray}
hence
\begin{eqnarray}
        \frac{\partial \mathcal{R}}{\partial x} = \partial^2_{tx} u(x,t) {+} \sum_{n=0}^{N} \partial_x\left[\partial_x (\partial_x + u)^n u\right]\,,\\
        \frac{\partial \mathcal{R}}{\partial t} = \partial^2_{t} u(x,t) {+} \sum_{n=0}^{N} \partial_t\left[\partial_x (\partial_x + u)^n u\right]\,,
\end{eqnarray}
The total loss function is now defined as:
\begin{eqnarray}
    L=w_{e} L_{e}+w_{d}L_{d}+w_g L_g\,, \label{loss_tot_grad}
\end{eqnarray}
where $w_g$ is a weight parameter that determines the significance of the gradient constraint \eqref{grad_constr} in the optimization procedure. Numerical experiments performed in \cite{Yu2022}, showed that there is an improvement in accuracy for $w_g\sim 10^{-2}$, while for $w_g\sim 1$ gPINNs underperform compared to regular PINNs.

\subsection{The cl-gPINN algorithm}
To impose the conservation of the quantity $C$ given by \eqref{conserved_quantity}, one should be able to compute $\phi(x,t)$ from $u(x,t)$ and then form the following loss term:
\begin{eqnarray}
    L_{c} = \frac{1}{M_{\mathcal{C}}} \sum_{j=1}^{M_{\mathcal{C}t}}\left[\sum_{i=0}^{M_{\mathcal{C}x}}\phi(x_i,t_j) - \sum_{i=0}^{M_{\mathcal{C}x}}\phi(x_i,t_0)\right]^2 \,, \label{Lc}
\end{eqnarray}
where $M_\mathcal{C} = M_{\mathcal{C}x}\times M_{\mathcal{C}t}$ is the number of points used for the evaluation of $L_C$ and 
\begin{eqnarray}
    \phi(x_i,t_j) = \exp\sum_{k=0}^i u(x_k,t_j)\Delta x\,,\quad i=1,...,M_{\mathcal{C}x}\,,\;j=0,...,M_{\mathcal{C}t}\,,
\end{eqnarray}
where $\Delta x = x_{i+1}-x_i$. The total  cl-gPINN loss is 
\begin{eqnarray}
        L=w_{e} L_{e}+w_{d}L_{d}+w_g L_g + w_c L_c\,, \label{loss_tot_cl+grad}
\end{eqnarray}
where $w_c$ is the weight of the conservation law constraint.

\subsection{Data generation}
\label{subsec_3.3}
The loss function \eqref{loss_tot} consists of two terms, each calculated on two different sets of points. The set of collocation points $\mathcal{P}$, used for imposing the PDE constraint \eqref{loss_eq}, is generated by perturbing a structured grid using a random number generator from a normal distribution. In contrast, the set of points $\mathcal{T}$, where measurements are available and the data constraint \eqref{loss_dat} is applied, is generated using a structured grid with equidistant nodes.

Inference of the parametric values $\alpha_1,...,\alpha_{N}$ requires the use of ``experimental'' data. These data are the values of the soliton amplitude at the points  $\mathcal{T}_i\in\mathcal{T}$. The values are obtained by perturbing an exact solution with a small Gaussian noise. That is 
\begin{eqnarray}
    u_{dat}(x_i,t_i) = u_{exact}(x_i,t_i) + Gauss(\mu,\sigma)\,, \label{u_dat}
\end{eqnarray}
where $u_{exact}$ is given by one of the classes of the exact solutions constructed in this paper, $(x_i,t_i)\in \mathcal{T}$ and $Gauss(\mu,\sigma)$ is the Gaussian noise term with $\mu=0$ being the mean of the normal distribution and $\sigma$ is the variance of the normal distribution associated with measurement errors and uncertainties. 

\subsection{Network architecture and training procedure}

We train the neural network by minimizing loss function \eqref{loss_tot} using the Adam optimizer \cite{Kingma2017} and variable learning  rate  using  {the 1-cycle learning rate policy \cite{Smith2018}}. The neural network weights are initialized using the uniform Xavier (or Glorot) initialization method described in \cite{Glorot2010}. The neural networks used in this study are multilayer perceptrons, i.e. feed-forward, fully-connected DNNs. The networks consist of an input layer of two neurons, 4 hidden layers with 64 neurons per layer and an output neuron that gives the value of the unknown function $u$ for given input $(x_i,t_i)$. We have considered two distinct activation functions for our networks, namely the sigmoid and the tanh, which have similar performance. It is important to note that no neural architecture search (NAS) method, such as Bayesian Optimization, was used to select this particular architecture. Instead, the number of hidden layers and neurons was determined empirically. The layers and neurons were gradually increased, with performance being tracked until the network achieved satisfactory results.

In each epoch  the neural network $u_{net}$ is fed with two sets of data points, the data $\mathcal{P}\in D = [x_{min},x_{max}]\times[t_{min}, t_{max}]\subset \mathbb{R}^2$ which are the randomly selected collocation points used for the evaluation of $L_{e}$ given by \eqref{loss_eq} and the data $\mathcal{T}\in D = [x_{min},x_{max}]\times[t_{min}, t_{max}]\subset \mathbb{R}^2$ which are points where measurements for the wave amplitude $u$ are available. These points along with the corresponding measurements $u_{dat}$ are used for the evaluation of $L_{d}$ through \eqref{loss_dat}. The network architecture and training procedure are summarized in Fig.~\ref{fig_train}.

\section{Numerical experiments and results}
\label{sec_IV}

We consider two different cases of soliton interactions: three-soliton fusion and three-soliton fission described by \eqref{generic_Burgers'_hierarchy} with $N=3$. Parametric inference is in principle possible for any member of the generic Burgers' hierarchy \eqref{generic_Burgers'_hierarchy}. However, we should note that the actual values of  the  parameters $\alpha_n$, $n=1,2,...,N$ used for the generation of the data values $u_{dat}$ may not be retrieved   if the number of  the interacting solitons is smaller than $N$. In this   case the PINN algorithm may  identify  an alternative set of parameters $\alpha_n$, $n=1,2,...,N$ that produces the same solitary dynamics. To understand this let us suppose that the wavenumbers of the interacting solitary structures are known and that their number is smaller than $N$. Then the number of propagation velocities is also smaller than the number of the unknown parameters $\alpha_n$ and the set of dispersion relations becomes an underdetermined set of linear equations for $\alpha_n$. This underdetermined system can be satisfied by different sets of parameters. Hence, in order to study three-soliton structures we will truncate the generic hierarchy to $N=3$ and consider $\alpha_0=0$ as the advection equation is  not important for our study.

\subsection{Three-soliton fusion}
\label{subsec_4.1}
A three-soliton solution is obtained from \eqref{gen_soliton_1} with $K=3$. We use this class of solutions to produce the values $u_{dat}$ at points $\mathcal{T}_i$, through Eq. \eqref{u_dat} with parameter values $\alpha_3 = -0.5$, $\alpha_2 = 0.9$,   $\alpha_1 = -0.3$, $k_1 = 0.9$,     $k_2 = -0.9$, $k_3 = 1.6$,  {$\phi_0 = 1.0$}, $\xi_1=\xi_2=\xi_3=0$,  to obtain three-soliton fusion at $(x=0,t=0)$.    We set the variance parameter $\sigma$ in \eqref{u_dat} equal to $0.03$.

\begin{figure}[h!]
    \centering
    \includegraphics[scale=0.6]{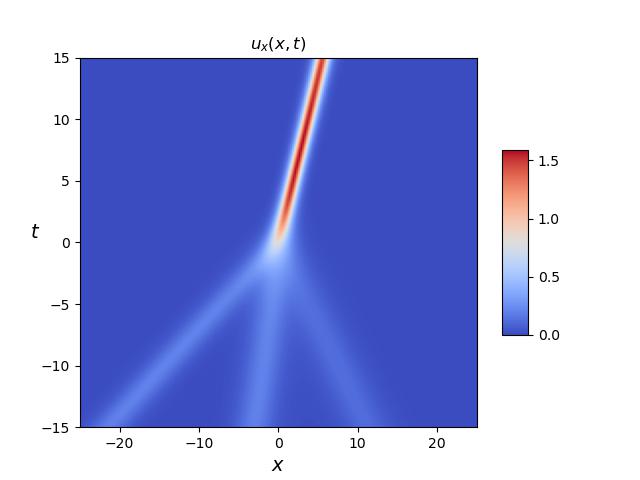}
    \caption{The derivative $u_x=\partial_x u$ of the neural network solution $u_{net}$ computed by the semi-supervised PINN algorithm, in the spatio-temporal domain $D=[-25,25]\times[-15,15]$ showing the fusion of three-solitons.}
    \label{fig_ux_2D_fusion}
\end{figure}

\begin{figure}[h!]
    \centering
    \includegraphics[scale=0.55]{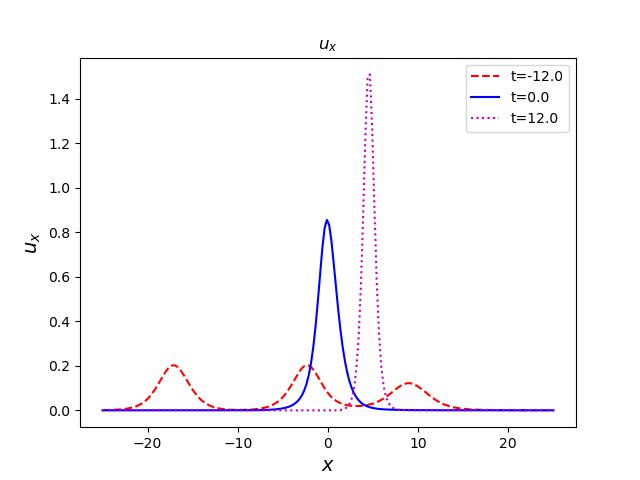}
    \caption{Three snapshots of $u_x=\partial_x u$ at times $t=-12$ (before fusion), $t=0.0$ (during fusion) and $t=12.0$ (after fusion).}
    \label{fig_snapshots_fusion}
\end{figure}

\begin{figure}[h!]
    \centering
    \includegraphics[scale=0.21]{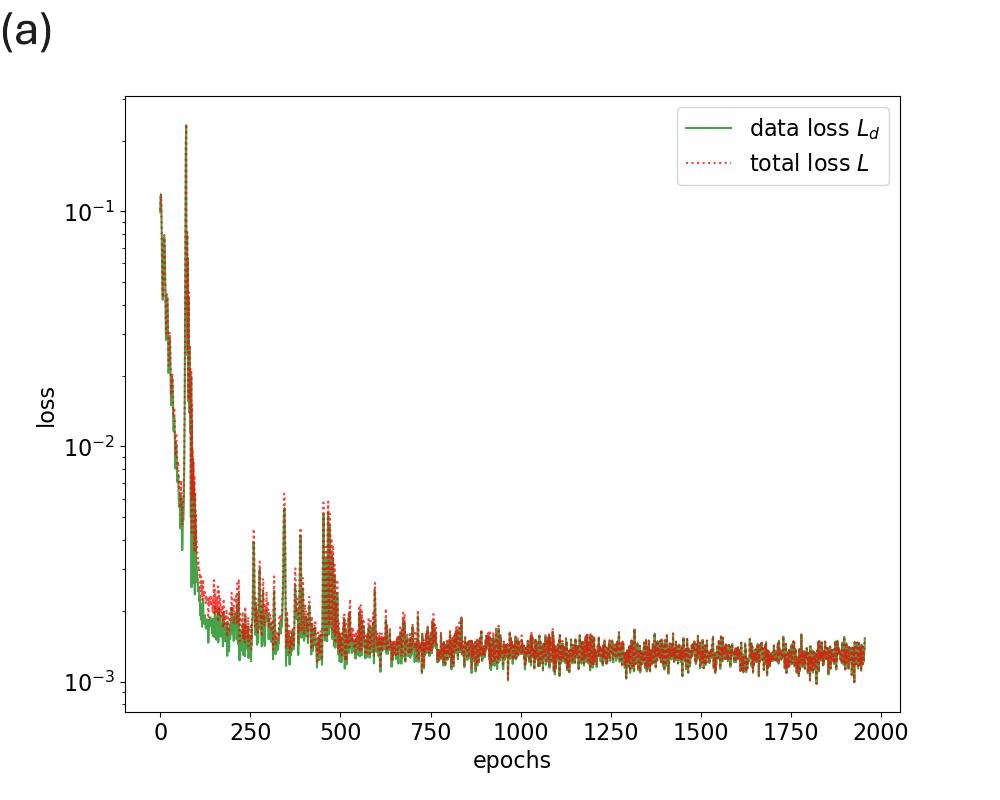}
    \includegraphics[scale=0.21]{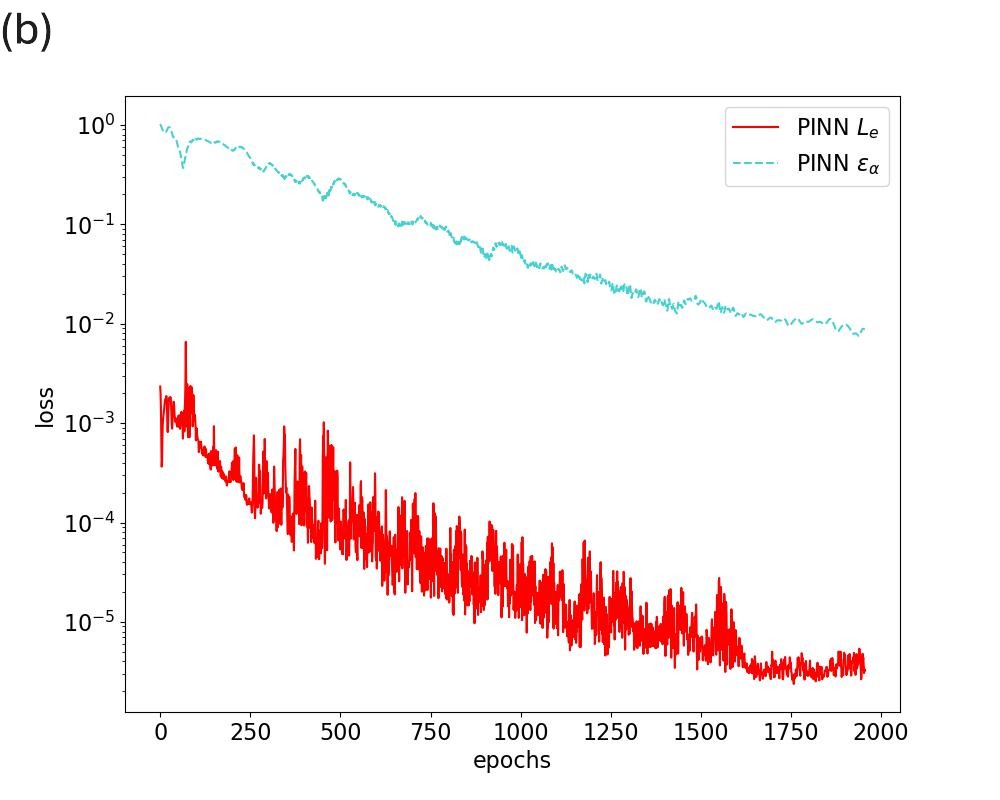}
    \caption{Training history for the soliton fusion numerical experiment. Panel (a) shows the total loss $L$ and the data loss $L_d$ plotted versus the training epochs. The data loss $L_d$ is constrained by the variance of noise used for data generation and cannot acquire smaller values. In (b) the equation loss $L_e$ and the alpha error $\epsilon_\alpha$, as defined in \eqref{alpha_error}, are shown. The equation loss $L_e$ decreases to the order of $10^{-5}$ after 1500 epochs, indicating that the data loss dominates the total loss function.}
    \label{fig_error_hist_fus}
\end{figure}

For this experiment we considered 800 collocation points $\mathcal{P}_i$ for the computation of $L_e$ and a coarser grid of $450$ points $\mathcal{T}_i$ with available data $u_{dat}(x_i,t_i)$. The computational domain is the rectangle $D=[-25,25]\times[-15,15]$. The maximum learning rate was set to $0.05$ and the network was trained for 2000 epochs { employing the 1-cycle learning rate policy \cite{Smith2018}}. Figure \ref{fig_ux_2D_fusion} displays the partial derivative of $u(x,t)$ with respect to $x$ on the spatio-temporal computational domain, showing three-solitons fusing at $(x=0,t=0)$ to form a single soliton for $t>0$. Three snapshots of this interaction are given in Fig. \ref{fig_snapshots_fusion} at times $t=-12.0$, $t=0.0$ and $t=12.0$.

In Fig. \ref{fig_error_hist_fus} we have plotted the training history for the fusion interaction.  The alpha error in \ref{fig_error_hist_fus} is a measure of the accuracy of the parametric inference and is defined as:
\begin{eqnarray}
    \epsilon_\alpha = \frac{1}{3} \sum_{i=1}^3 \frac{|\alpha_i-\bar{\alpha}_i|}{|\bar{\alpha}_i|}\,, \label{alpha_error}
\end{eqnarray}
where $\bar{\alpha}_i\,,\; i=1,2,3$ are the exact parametric values used for the generation of data through \eqref{u_dat} and \eqref{gen_soliton_1}.

 The equation loss $L_e$ decreases to the order of $10^{-5}$ after 1500 epochs, indicating that the data loss dominates the total loss function. This is because the data loss is constrained by the variance of the noise used for data generation, preventing it from falling below $10^{-3}$. Additionally, the alpha error stabilizes below $10^{-2}$, demonstrating successful parameter inference. { The precise minimum values of the training history metrics are summarized in table \ref{tab_1}.
}

\subsection{Three-Soliton fission}
\label{subsec_4.2}

For the fission of a solitary structure to 3 solitons at $(x=0,t=0)$ we set $\alpha_3 = 0.6$, $\alpha_2 = -0.9$,   $\alpha_1 = 0.2$, $k_1 = 0.9$,     $k_2 = -0.9$, $k_3 = 1.6$, { $\phi_0 = 1.0$}, $\xi_1=\xi_2=\xi_3=0$.

\begin{figure}[ht!]
    \centering
    \includegraphics[scale=0.6]{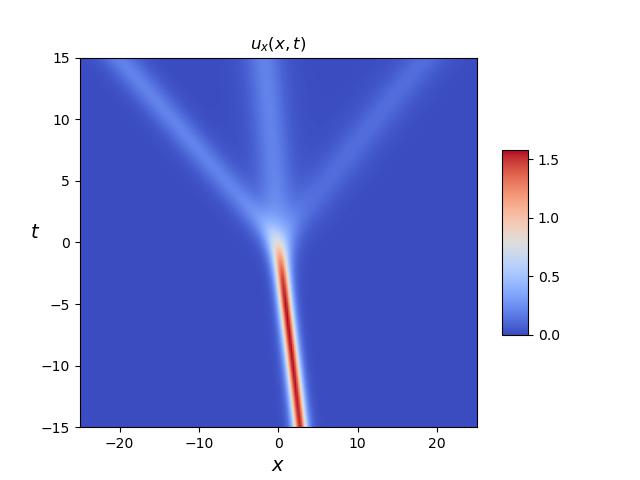}
    \caption{The derivative $u_x=\partial_x u$ of the neural network solution $u_{net}$ in the spatio-temporal domain $D=[-25,25]\times[-15,15]$ showing the fission of a solitary wave to three-solitons.}
    \label{fig_ux_2D_fission}
\end{figure}

\begin{figure}[ht!]
    \centering
    \includegraphics[scale=0.5]{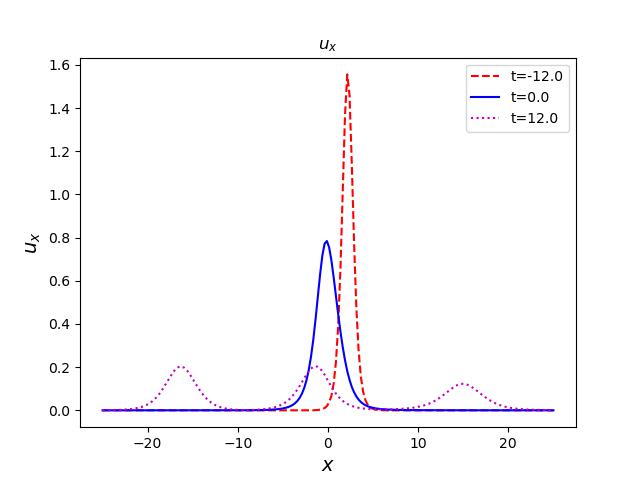}
    \caption{Three snapshots of $u_x=\partial_x u$ at times $t=-12$ (before fission), $t=0.0$ (during fission) and $t=12.0$ (after fission).}
    \label{fig_snapshots_fission}
\end{figure}

\begin{figure}[ht!]
    \centering
    \includegraphics[scale=0.215]{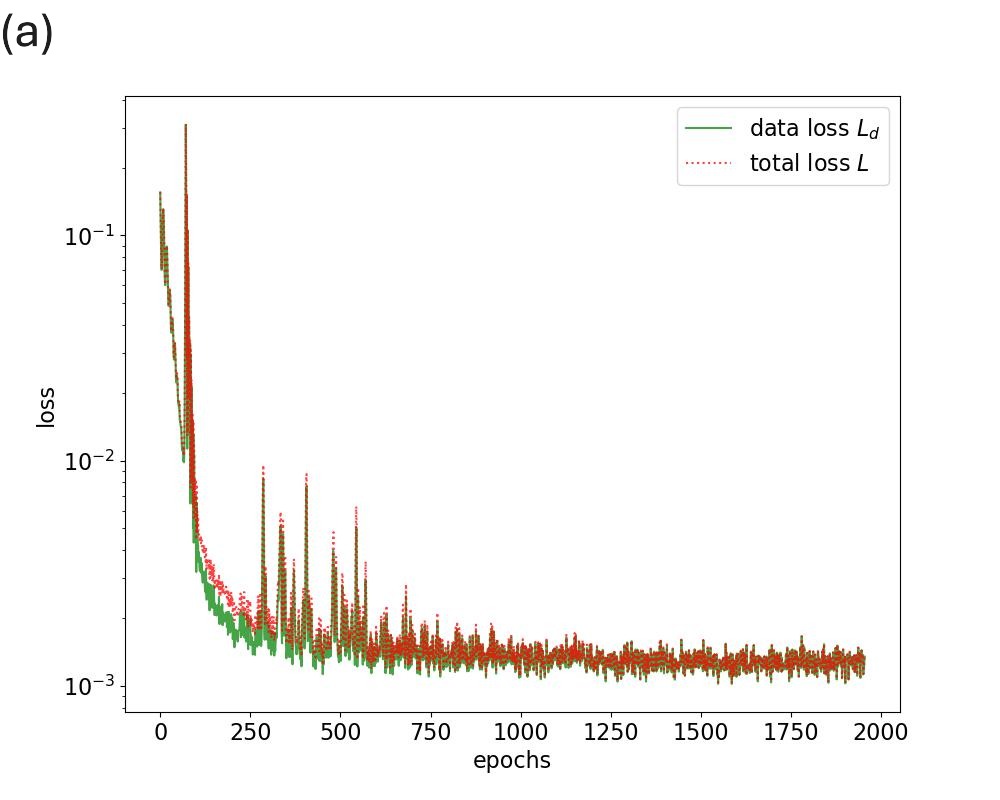}
    \includegraphics[scale=0.215]{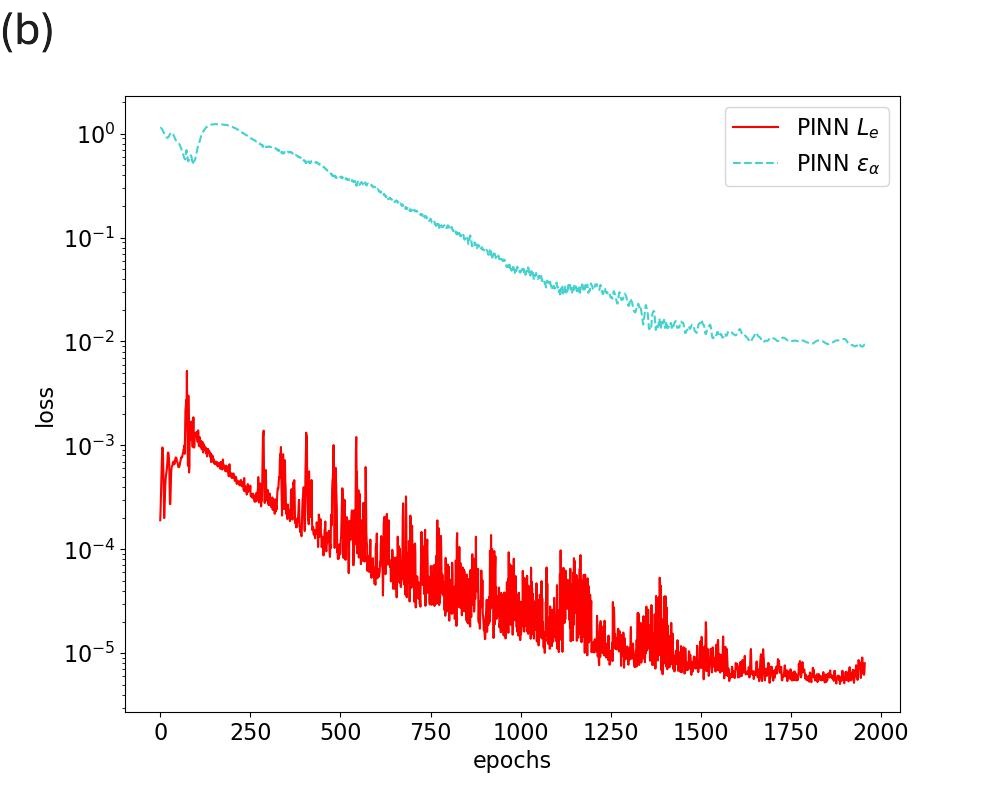}
    \caption{Training history for the soliton fission numerical experiment. Panel (a) shows the total loss $L$ and the data loss $L_d$ plotted versus the training epochs. In panel (b) the equation loss $L_e$ and the alpha error $\epsilon_\alpha$, are shown. The equation loss $L_e$ decreases to the order of $10^{-5}$ after 1500 epochs, demonstrating the effectiveness of the training.}
    \label{fig_error_hist_fis}
\end{figure}

Apart from this change of parameters, we considered  the same computational domain $D$, number of collocation and data points, variance for the generation of  $u_{dat}$ and the same training hyperparameters as in the previous case discussed in the subsection \ref{subsec_4.1}. The results of this simulation are presented in Figs. \ref{fig_ux_2D_fission}--\ref{fig_error_hist_fis}. In Fig. \ref{fig_ux_2D_fission} we show $\partial_x u(x,t)$ plotted on the domain $D$, while in \ref{fig_snapshots_fission} we have three snapshots, before, during and after the soliton fission.

In Fig. \ref{fig_error_hist_fis} we show the trainig history of this second simulation  which  indicates that the loss function terms $L_e$, $L_d$ and the total loss are minimized effectively, while $\epsilon_\alpha$ is of the order of $10^{-2}$ demonstrating that the parametric values have been successfully inferred.

\begin{table}[h!]
{
    \centering
  \begin{tabular}{lllll} 
    \toprule
    {case}&{$L $}& {$L_e$}& {$L_d$} & {$\epsilon_\alpha$}   \\
    \midrule
{fusion} & $1.308 \times 10^{-3}$ & $2.8\times10^{-6}$ &   $1.305\times 10^{-3}$ & $8.6\times 10^{-3}$ \\
{fission} & $1.147\times10^{-3}$  & $1.142\times10^{-5}$ & $5.3\times10^{-6}$ & $1.04\times 10^{-2}$\\
    \bottomrule
\end{tabular}
    \caption{Minimum values of the training history metrics for the 3-soliton fusion and fission cases described in sections \ref{subsec_4.1} and \ref{subsec_4.2}, respectively, using the baseline PINN approach.}
    \label{tab_1}
    }
\end{table}

\subsection{Convergence analysis}
We conducted a convergence analysis by increasing the number of collocation points, $\Mc_P$, and data points, $\Mc_T$ (collectively referred to as sample points), while measuring the total, equation, and data mean absolute errors, $L_1$, $L_{1e}$, and $L_{1d}$, respectively, to identify an optimal grid size that balances accuracy and computational efficiency. The mean absolute errors were evaluated on validation test points obtained by perturbing the positions of the training points. For this analysis, we used an equal number of points for the equation and the measured data, i.e. $\Mc_P=\Mc_T$. Additionally, we recorded the training time and calculated the alpha error, as defined in Eq. \eqref{alpha_error}, for various numbers of sample points.
\begin{figure}[ht!]
    \centering
    \includegraphics[width=0.475\linewidth]{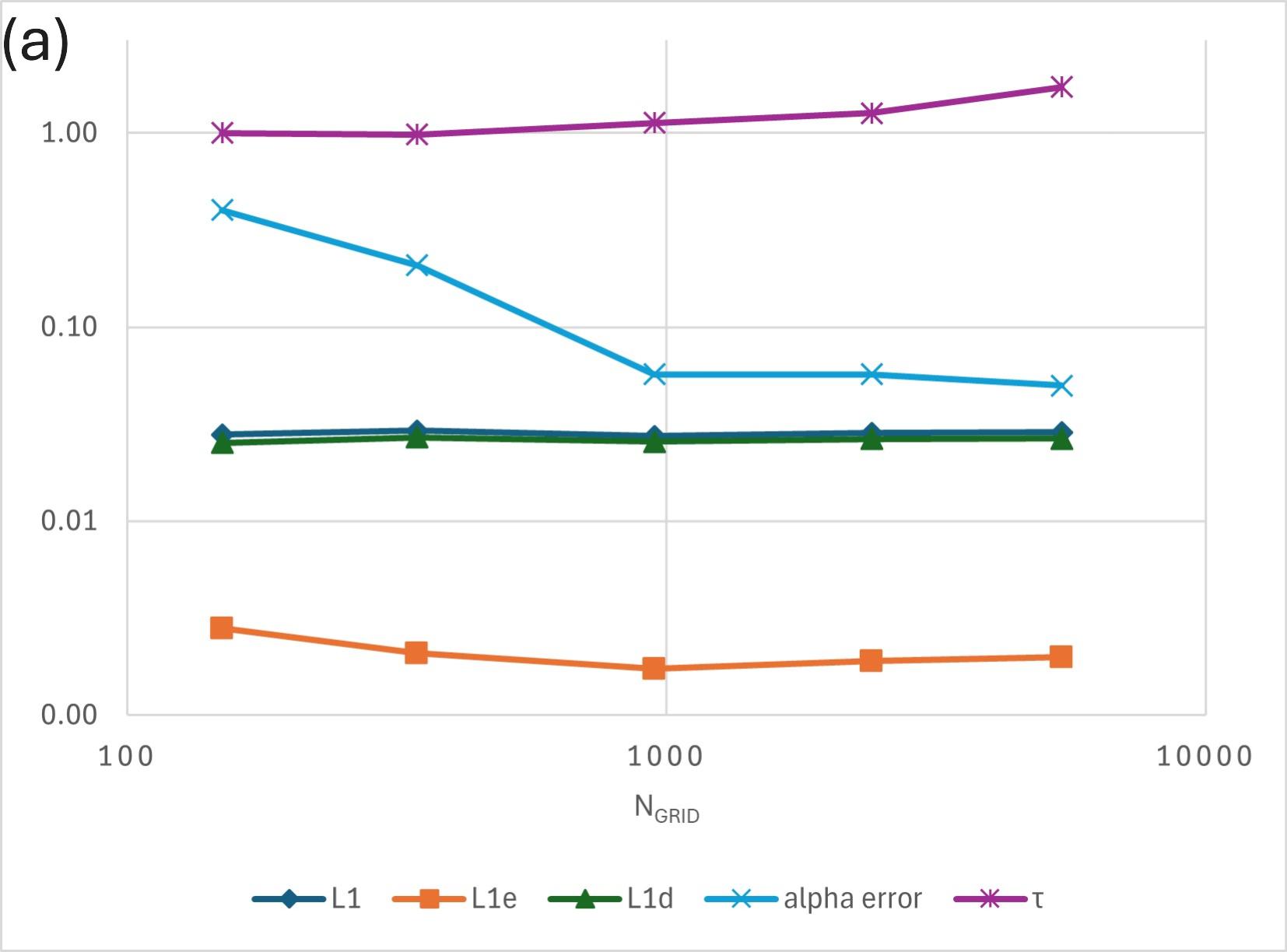}
    \includegraphics[width=0.475\linewidth]{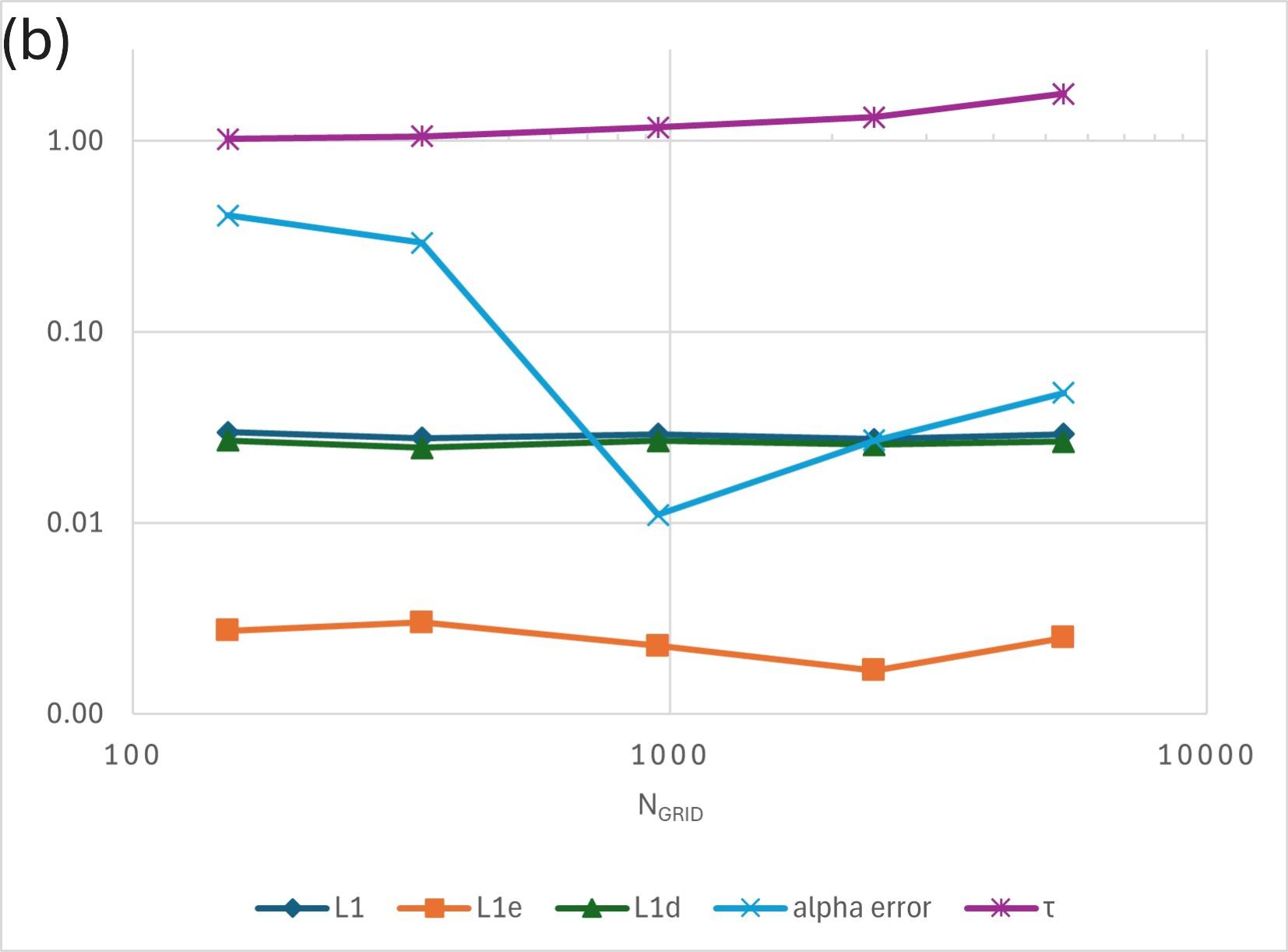}
    \caption{{The mean  absolute errors $L_1=L_{1e}+L_{1d}$, $L_{1e}=(1/N)\sum_{i=1}^N|\Rc_i|$, $L_{1d}=(1/N)\sum_{i=1}^N|u_{net}^i-u_{dat}^i|$ are plotted versus the grid size for the soliton fusion (a) and fission (b) cases. The alpha error and the slowdown factor $\tau=T_N/T_{coarse}$, are also shown in the same diagram. Here $T_N$ is the training time for the $N$-point grid and $T_{coarse}$ is the training time for the coarser grid. For this comparison we trained the network with a constant learning rate $\eta=0.01$. }}
    \label{fig_L1_fus-fis}
\end{figure}
The results summarized in Fig. \eqref{fig_L1_fus-fis} indicate that, for the soliton fusion case (a), the equation residual error is minimized at approximately $10^3$ sample points. The error $\epsilon_\alpha$ decreases rapidly as the number of sample points increases from $10^2$ to $10^3$; however, for finer grids, the accuracy gains become negligible or even reversed for $L_{1e}$, while the computational time increases significantly. Additionally, the total $L_1$ error is primarily dominated by the data mean  absolute error $L_{1d}$, which remains nearly constant across all grid sizes, due to the noise variance. 

For the soliton fission case (b) the total mean absolute error $L_1$ is again dominated by the data error while the equation residual error is minimized for grids with a number of points  between $10^3$ and $10^4$. Also, the alpha error $\epsilon_\alpha$ that characterizes parameter inference accuracy, is most effectively minimized for $\sim 10^3$ sample points but increases for finer grids. Thus, we conclude that for both the soliton fusion and fission cases, a grid size of approximately $\sim 10^3$ sample points achieves the best balance of accuracy and computational efficiency.

\subsection{Noise analysis}
\begin{figure}[ht!]
    \centering
    \includegraphics[width=0.475\linewidth]{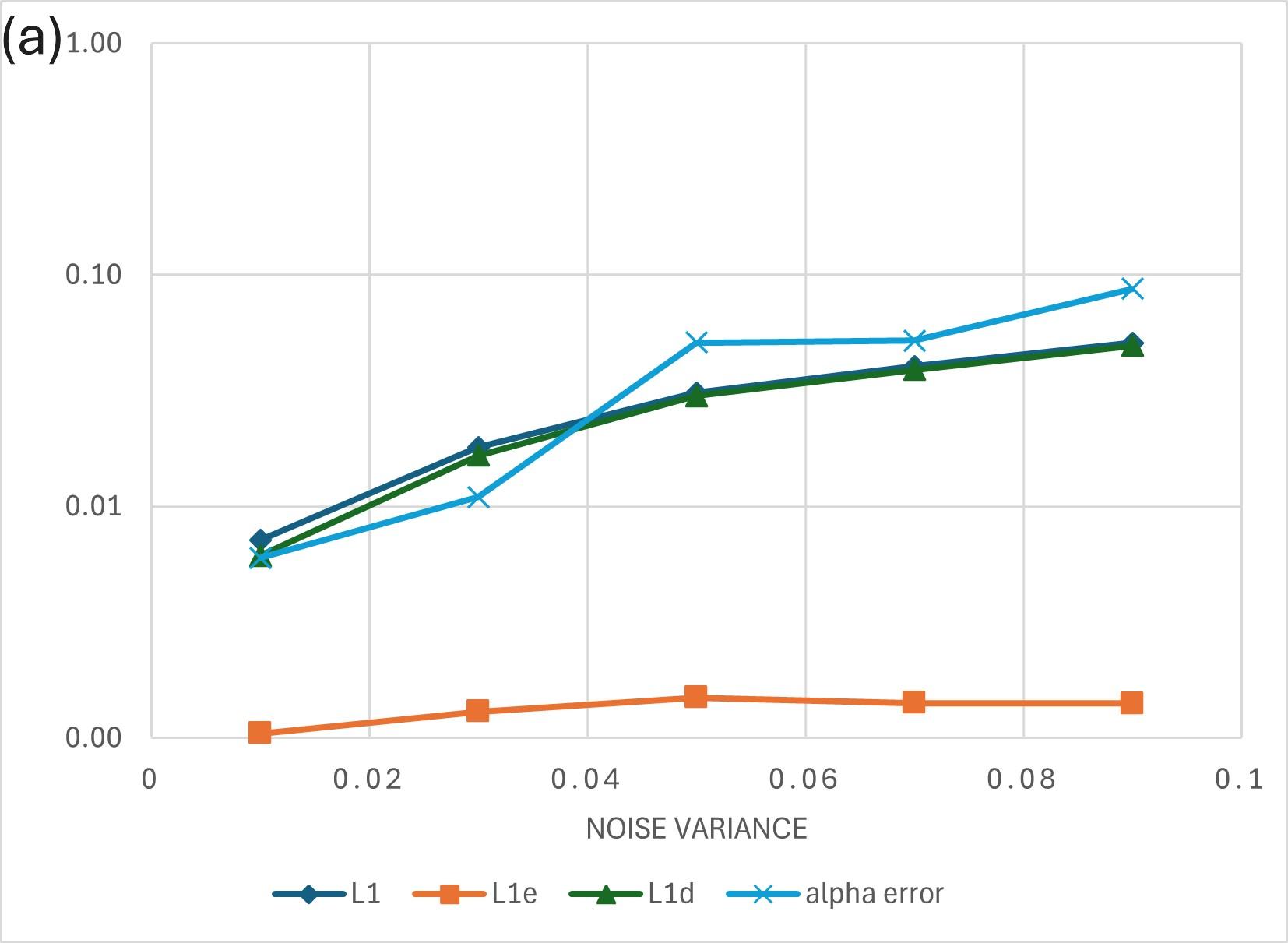}    \includegraphics[width=0.475\linewidth]{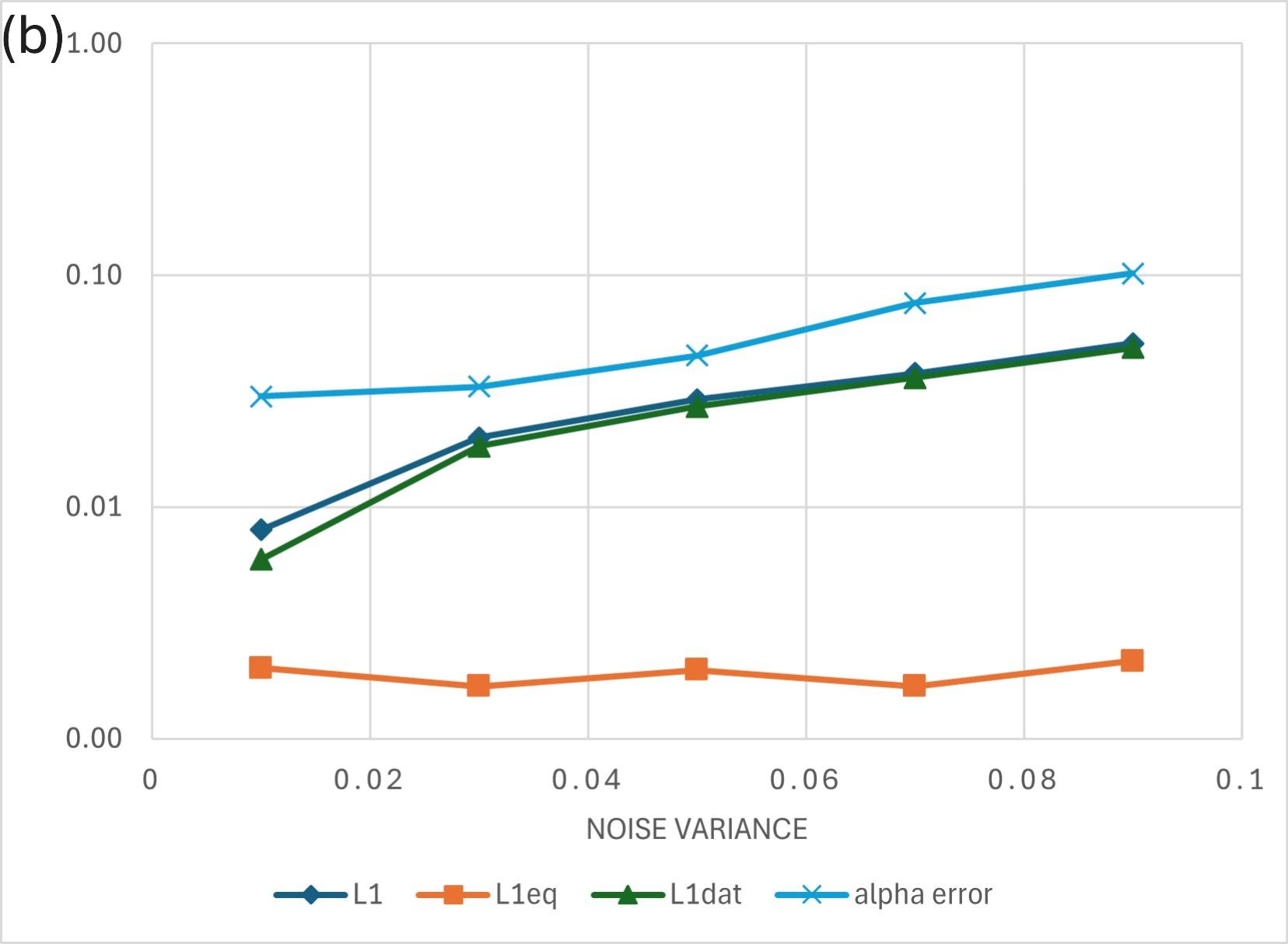}
    \caption{{The dependence of the  alpha error $\epsilon_\alpha$ and the total, equation and data mean  absolute errors on  the Gaussian noise variance $\sigma$ for the three soliton fusion (a) and fission (b) cases.}}
    \label{fig_noise_fus-fis}
\end{figure}

To estimate  the impact of the Gaussian noise on the accuracy  of the neural network solutions and parameter inference, we consider 5 different noise levels with variances from  $\sigma =0.01$ to $\sigma=0.09$. These experiments are performed using the baseline number of 800 collocation points for the equation loss and 450 points for the data loss term. The total, equation and data mean absolute errors $L_1$, $L_{1e}$ and $L_{1d}$ are measured for each noise level along with the alpha error $\epsilon_\alpha$. The results are summarized in Fig. \ref{fig_noise_fus-fis}. By these plots, it is evident that the alpha error $\epsilon_\alpha$ and the data mean average error $L_{1d}$ are increasing with noise variance both in the soliton fusion (a) and soliton fission case (b). This indicates that greater noise levels result in greater errors in the inferred parameters, i.e. the inferred parameters deviate more from their true values, which has also been observed in \cite{Wang2024a}. On the other hand the equation error ($L_{1e}$) remains nearly constant with small increase in the fusion case (a).
 
\begin{figure}[ht!]
    \centering
    \includegraphics[width=0.45\linewidth]{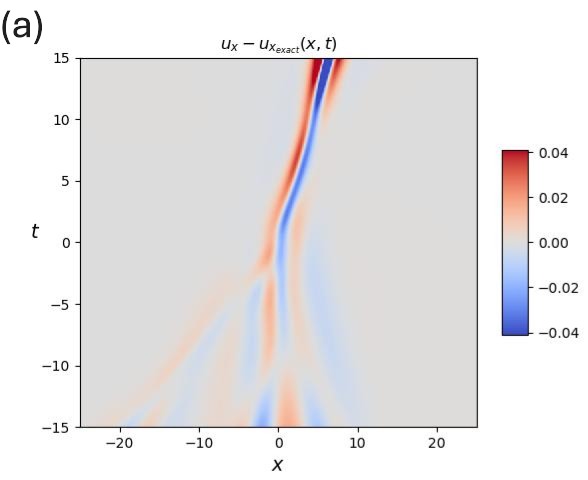}    \includegraphics[width=0.45\linewidth]{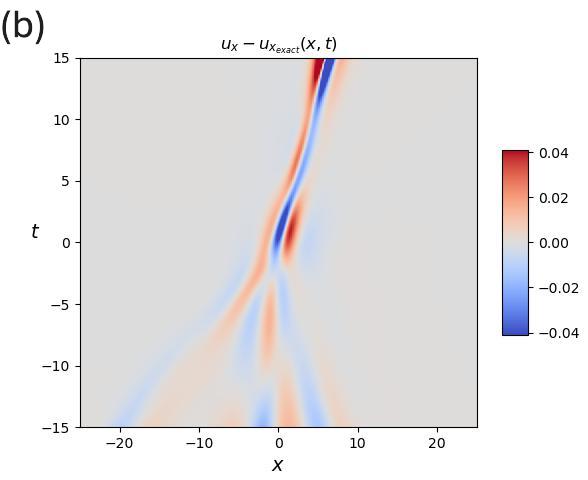}   \\\includegraphics[width=0.45\linewidth]{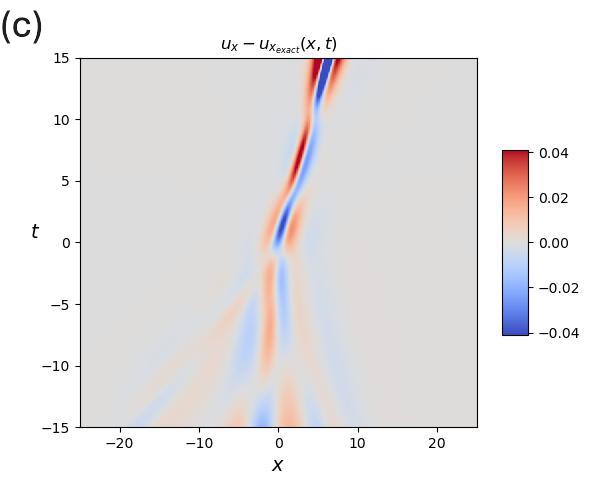} \includegraphics[width=0.45\linewidth]{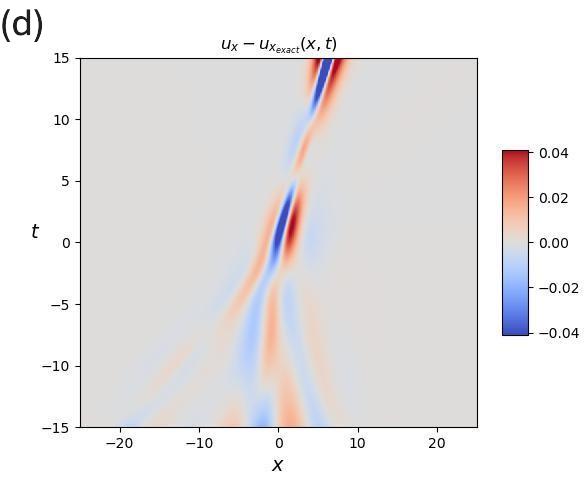}
    \caption{{Four different instances of the error of the DNN solution with noise variances $\sigma=0.03$ (a), $\sigma=0.05$ (b), $\sigma=0.07$ (c) and $\sigma=0.09$ (d) for soliton fusion.}}
    \label{fig_errors_fus}
\end{figure}
\begin{figure}[ht!]
    \centering
    \includegraphics[width=0.45\linewidth]{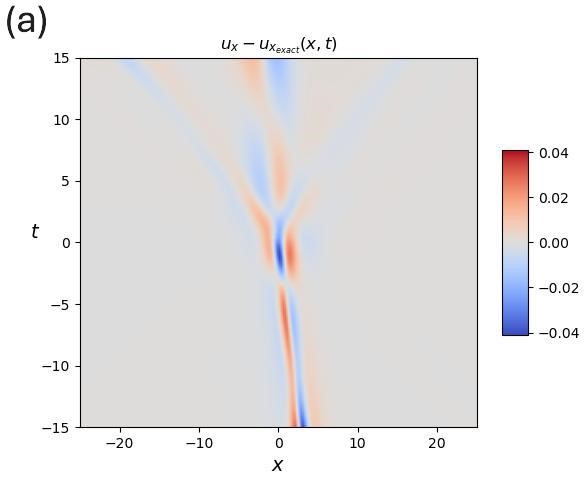}\includegraphics[width=0.45\linewidth]{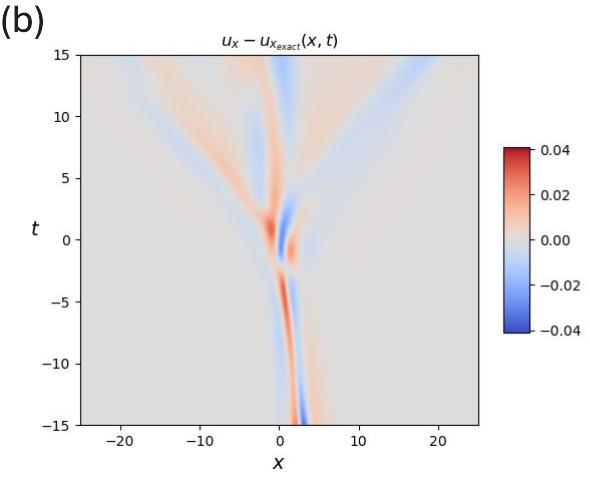}\\
    \includegraphics[width=0.45\linewidth]{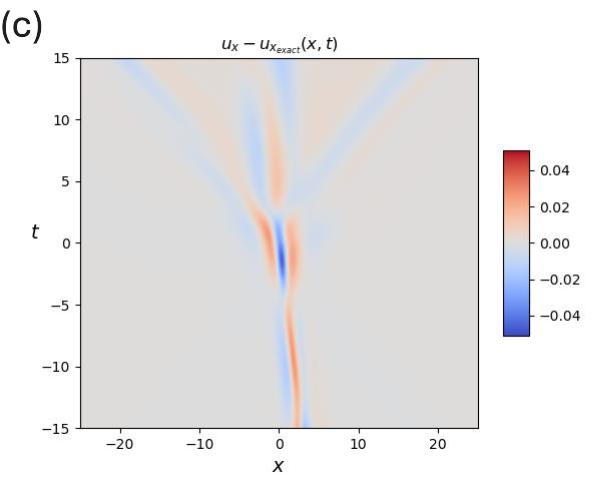}
    \includegraphics[width=0.45\linewidth]{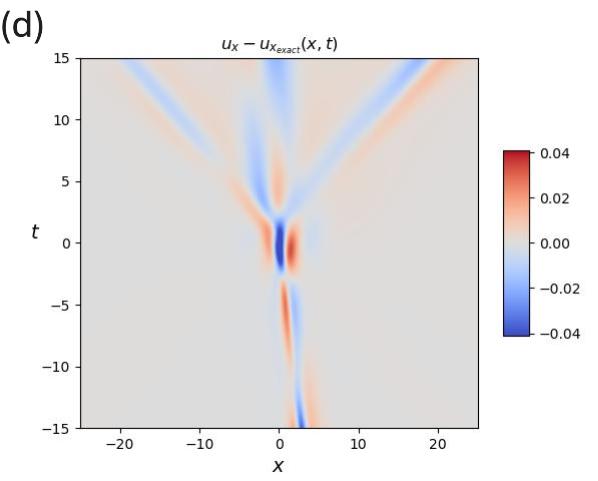}
    
    \caption{{Four different instances of the error of the DNN solution with noise variances $\sigma=0.03$ (a), $\sigma=0.05$ (b), $\sigma=0.07$ (c) and $\sigma=0.09$ (d) for soliton fission.}}
    \label{fig_errors_fiss}
\end{figure}

In Fig. \ref{fig_errors_fus}, we plot the difference $u_{net}(x,t) - u_{exact}(x,t)$, where $u_{exact}$ was used to generate the experimental data, for four noise variances: $\sigma = 0.03$, $\sigma = 0.05$, $\sigma=0.07$ and $\sigma = 0.09$, for the soliton fusion. This comparison demonstrates that the trained solutions approximate the exact solution well, with maximum differences on the order of $10^{-2}$. Although we could not confirm that increasing noise reduces error, as suggested in \cite{Wang2024a}, our results show that the approximation remains robust across the various levels of Gaussian noise considered in this study.
 
For the soliton fission, four different instances of the error of the DNN solution with respect  to the exact  solution, with noise variances $\sigma=0.03$, $\sigma=0.05$, $\sigma=0.07$ and $\sigma=0.09$, are provided in Fig. \ref{fig_errors_fiss} . As  in  the previous case, the  error  is on the order of $10^{-2}$ and remains robust across the various levels of Gaussian noise.

\subsection{Solution prediction at future times}
To assess the model's ability to predict unseen data at future times, we conducted additional numerical experiments for both the soliton fusion and fission cases. In these experiments, the total time interval $[t_{min}, t_{max}]$ is divided into two sub-intervals: $[t_{min}, t_0]$, representing the interval with observational data available up to the present time $t_0$, and $[t_0, t_{max}]$, representing the unseen interval. The model is trained on the available data within $[t_{min}, t_0]$, and its ability to predict the soliton evolution in the unseen interval $[t_0, t_{max}]$ is assessed by evaluating the mean absolute error $L_{1p}$ between the predicted and exact solutions in this interval. We also evaluated the errors $L_{1e}$, $L_{1d}$, and $\epsilon_\alpha$ as previously done in subsection \ref{subsec_4.1}. We observed that for $L_{1p} > 0.1$, the predicted solution becomes morphologically inconsistent with the true behavior. Therefore, we set a threshold of $0.1$ as the limit for how far into the future we can extend our predictions. For both the soliton fusion and fission cases, this occurs when $t_0 > 0$, i.e., after the fusion/fission event. Moreover, the error increases as $t_0$ decreases, which corresponds to an increase in the ratio $(t_{max} - t_0)/(t_0 - t_{min})$.

\begin{figure}[ht!]
    \centering
    \includegraphics[width=0.475\linewidth]{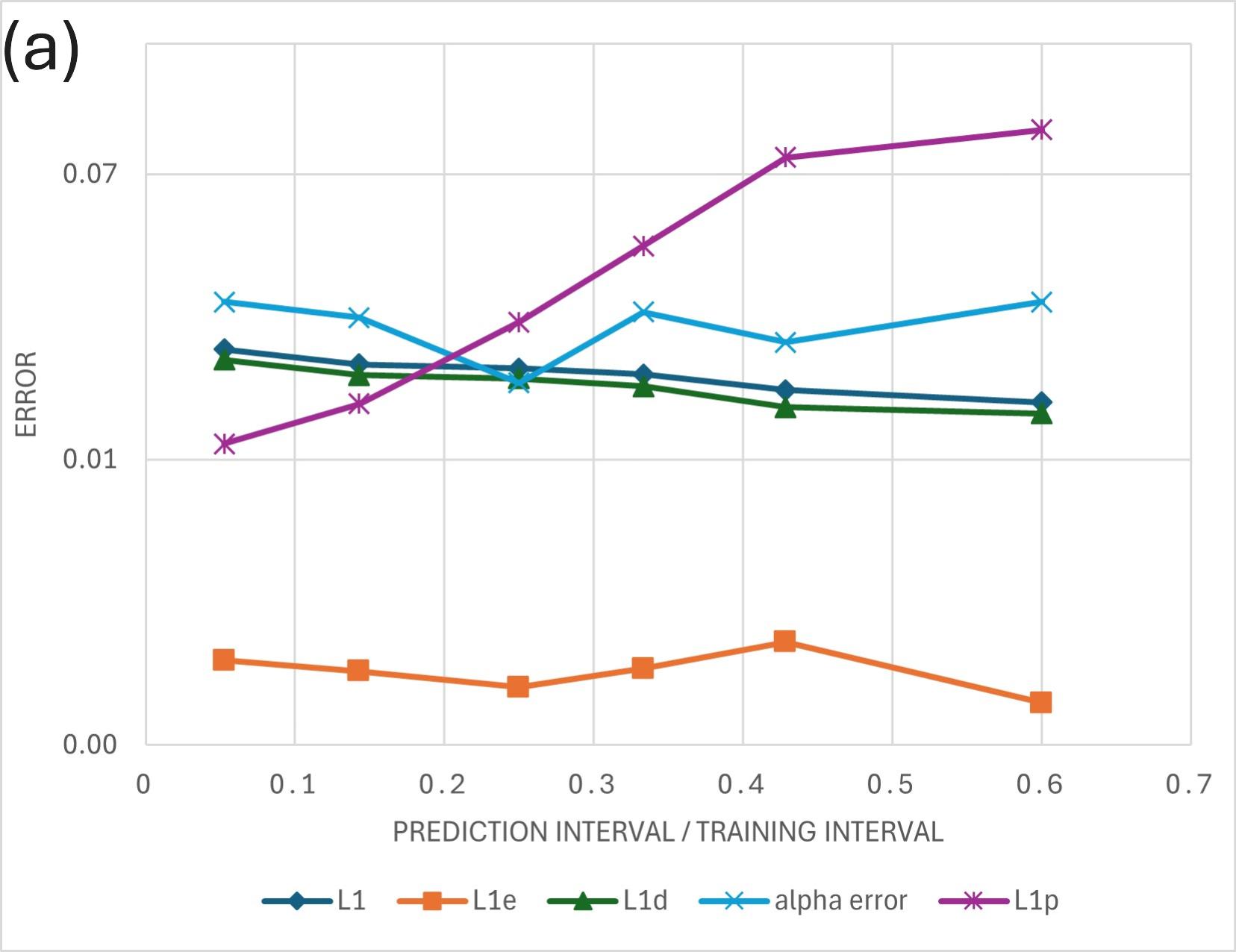}
    \includegraphics[width=0.475\linewidth]{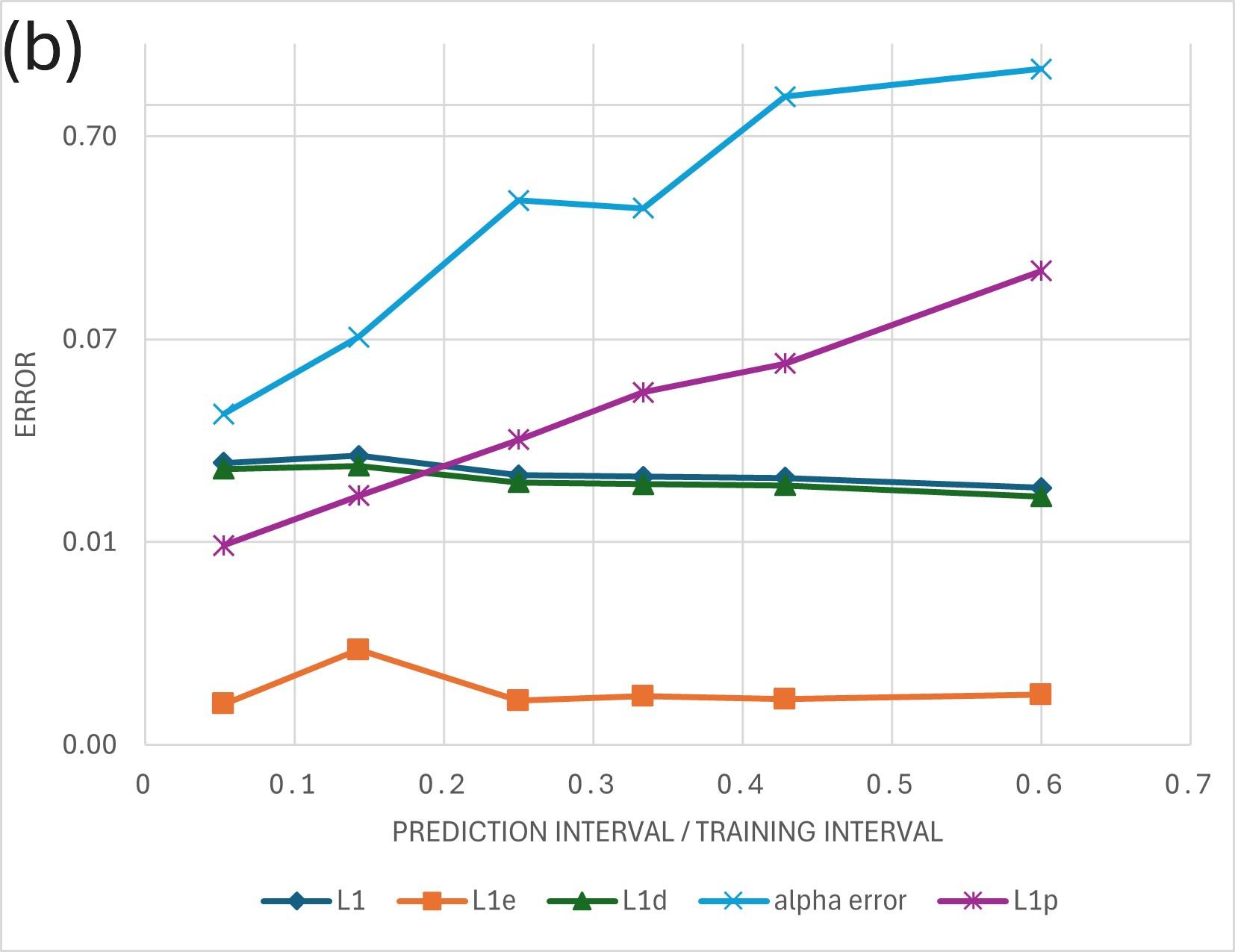}
    \caption{Accuracy metrics for the soliton fusion (a) and fission (b) cases, evaluated over a time interval divided into training and prediction sub-intervals.}
    \label{fig_L1_pred}
\end{figure}

In Fig. \ref{fig_L1_pred}, we present accuracy metrics for the soliton fusion (a) and fission (b) cases, evaluated over a time interval divided into training and prediction sub-intervals. In both cases, the mean average $L_{1p}$ error increases as the ratio of the prediction interval to the training interval increases. For the soliton fusion and fission cases, the accuracy of the predicted solution remains within acceptable bounds when $(t_{max} - t_0)/(t_0 - t_{min}) < 0.18$ and $(t_{max} - t_0)/(t_0 - t_{min}) < 0.2$, respectively, with $L_{1p} < L_{1d}$ in these ranges.

We also observe that, in the soliton fusion case, parameter inference is largely unaffected by this ratio as long as it remains below 1. In contrast, this is not the case for the soliton fission scenario. This difference arises because parameter inference requires data from three distinct modes with different speeds. In soliton fusion, data from three converging modes are available even when $t_0 = 0$. However, for soliton fission, to ``see'' the diverging waves, we need $t_0 > 0$, and the available data increase as $t_0$ increases.

\begin{figure}[ht!]
    \centering
    \includegraphics[width=0.475\linewidth]{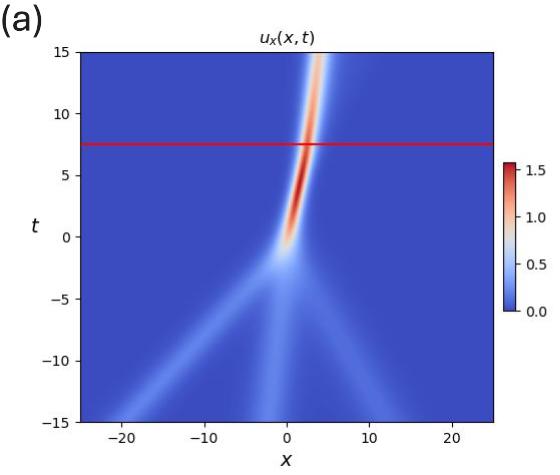}
    \includegraphics[width=0.475\linewidth]{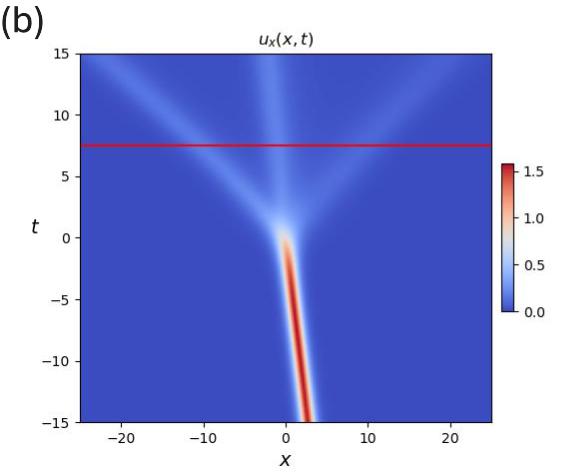}
    \caption{The trained and predicted parts of the DNN solutions for the soliton fusion (a) and fission (b) experiments. The horizontal red line corresponds to $t_0$, i.e. is the boundary between the training and the prediction intervals.}
    \label{fig_pred}
\end{figure}

Additionally, in Fig. \ref{fig_pred}, we show the trained and predicted solutions in the same density plots for $t_0 = 7.5$, corresponding to $(t_{max} - t_0)/(t_0 - t_{min}) \approx 0.33$.
The trained network can reconstruct the solution at future times with small deviations from the true behavior (see Figs. \ref{fig_ux_2D_fusion} and \ref{fig_ux_2D_fission}) which are more pronounced for the soliton fusion.

\subsection{The cl-gPINN enhancement}
The  cl-gPINN algorithm consists in determining the $u_{net}$ internal parameters and the PDE parameters $\alpha_n$, $n=1,2,3$, by minimizing the loss function \eqref{loss_tot_cl+grad}. After numerous experiments we concluded that cl-gPINN can  improve training   accuracy in specific  cases and after  a fine tunning  of the weights $w_{g}$ and  $w_c$. We have corroborated  the reported  \cite{Yu2022} tendency  of gPINNs  to  provide  enhanced accuracy   for $w_g\sim   10^{-2}$. In addition, $w_{c}$ should be adjusted in such a way so that the term $w_c L_c$ in \eqref{loss_tot_cl+grad} does not dominate the other terms. Even if this care is taken, the cl-gPINN method improves accuracy compared to the regular PINNs only if the learning rate is comparatively large throughout the entire training. In most cases we observed significant improvement for learning rates $\geq0.1$, as imposing the grad constraint and the conservation law constraint seemed to stabilize the equation loss term and improved the parameter inference. However, for small or dynamically adapting learning rates, this advantage of the cl-gPINN approach seems to be diminished or reversed. In addition to that, both the imposition of the gradient loss term and the conservation law term increase the computational cost since the former requires the computation of higher order derivatives while the latter requires the intermediate numerical estimation of the auxiliary variable $\phi(x,t)$. 

To provide a couple of comparisons considering soliton fusion and fission, we have implemented the following practice: initially we used the regular PINN method minimizing the loss \eqref{loss_tot}. The network parameters were initialized with the uniform Xavier initialization method. This initial state of the network was saved in order to initialize the network again for a second run minimizing \eqref{loss_tot_grad}, a third run with the conservation law constraint replacing the gradient term and a fourth run minimizing \eqref{loss_tot_cl+grad}. It is important for the reliability of this comparison that in all cases the same initial state was used. Also, in order to avoid any modulation due to the distribution of the collocation points we considered structured grids of equidistant collocation points. Three such grids where constructed, one for calculation the PDE loss term $L_e$, a second for calculating the data loss term $L_d$ and a third for imposing the conservation law $L_c$, with $25\times15=375$, $20\times 10 =200$ and $40\times10=400$ points, respectively. We considered three-soliton fusion and fission and we trained the network for 2000 epochs with a static learning rate. The learning rate was set to a fixed value large enough to cause the baseline PINN approach to fail in parametric inference (i.e. $\epsilon_\alpha>1$).  We also changed the architecture of the network, which now has 2 hidden layers with 64 neurons per layer.

\begin{figure}[ht!]
    \centering
    \includegraphics[scale=0.21]{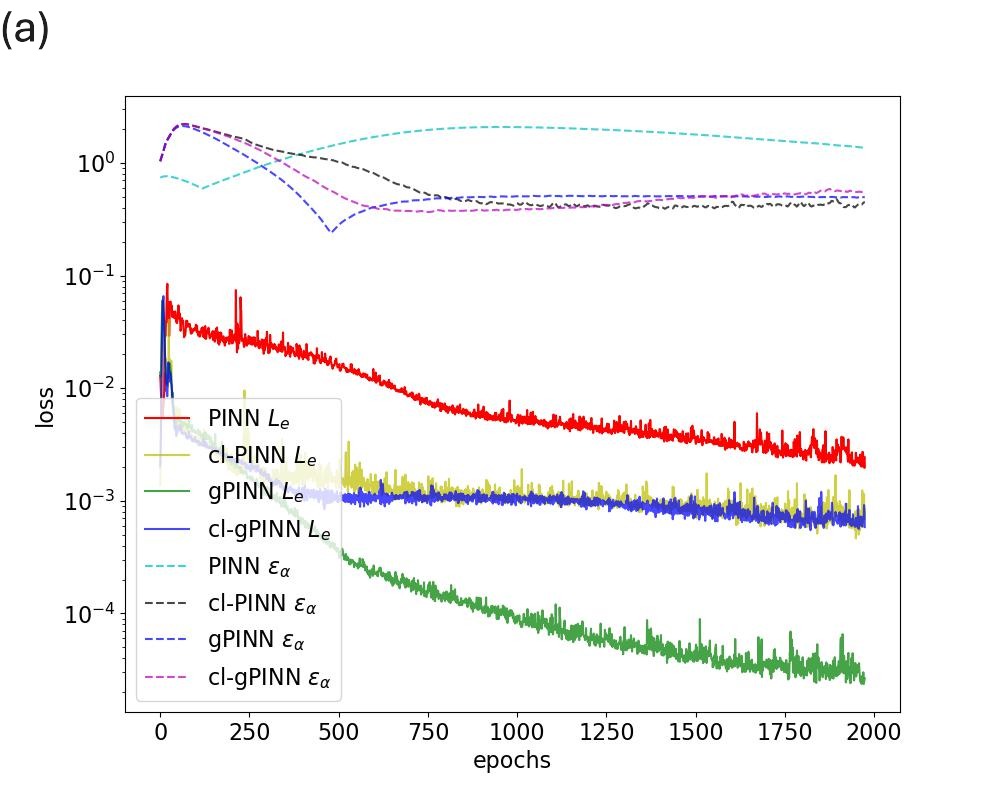}
        \includegraphics[scale=0.21]{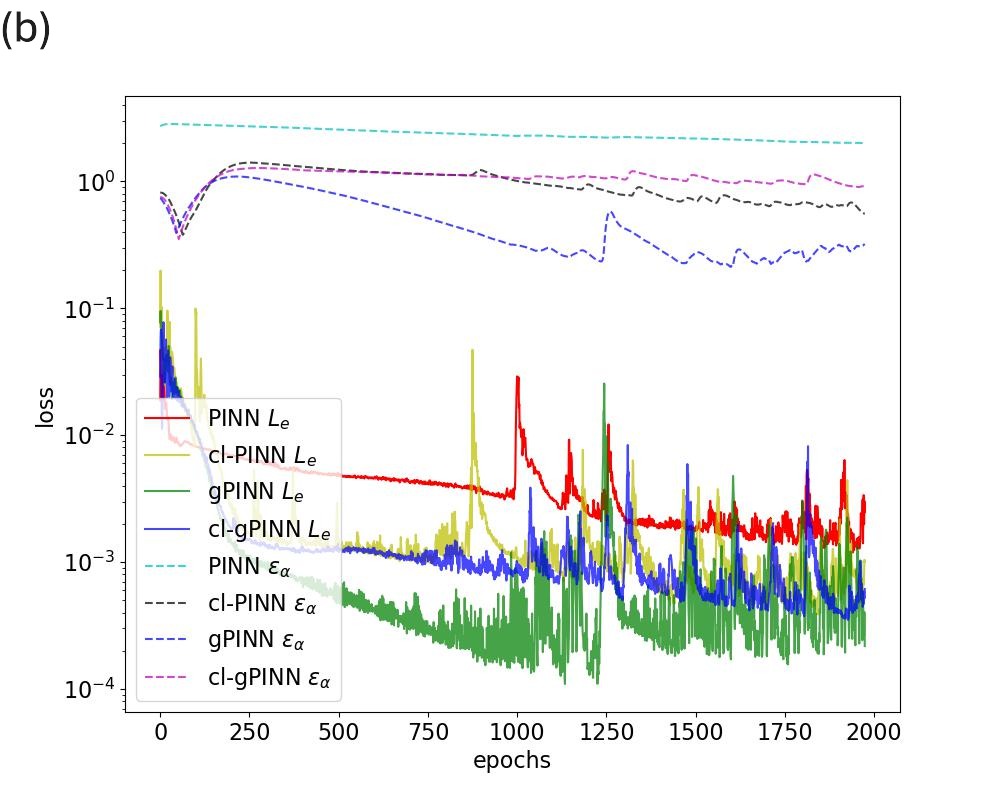}
    \caption{Training history for the soliton  fusion (a) and soliton fission (b) numerical experiments, demonstrating the enhancement provided by the gPINN and cl-gPINN algorithms compared to regular PINNs. The equation loss is minimized more effectively in both cases, resulting in an accuracy improvement. Parameter inference is possible only with the enhanced algorithms, as PINNs are unable to infer parametric values in this training setting.}
    \label{fig_hist_enh_fus-fis}
\end{figure}

Figure (a) in \ref{fig_hist_enh_fus-fis} shows the training history for the soliton fusion numerical experiment with parameter values $\alpha_3 = -0.6$, $\alpha_2 = 0.9$,   $\alpha_1 = 0.3$, $k_1 = 0.9$,     $k_2 = -0.9$, $k_3 = 1.6$,  {$\phi_0 = 1.0$, $\xi_1=\xi_2=\xi_3=0$. The numerical value of $L_e$ in each case during the best training epoch (with minimum $L$) and the final value of $\epsilon_\alpha$ are summarized in table \ref{tab_2}. For this training setting—with coarse grids of points $\Pc$ and $\Tc$ and a comparatively large, constant learning rate—the gPINN and cl-gPINN algorithms provide a significant improvement over regular PINNs. The equation loss $L_e$ is minimized more effectively, leading to greater accuracy. Parameter inference is also improved: while infeasible with the standard PINN approach, it becomes more feasible using gPINNs, cl-PINNs, and cl-gPINNs. However, this behavior is fragile and heavily dependent on the initialization of network parameters. In some cases, the PINN algorithm effectively minimizes both $L_e$ and $\epsilon_\alpha$, rendering the enhancements of gPINN, cl-PINN, and cl-gPINN negligible or even nonexistent. In such cases, by increasing the learning rate we can make the PINN algorithm to miss the global minimum of $L$ and overshoot, while the enhanced algorithms provide increased accuracy.}

In figure (b) of \ref{fig_hist_enh_fus-fis} are given the corresponding diagrams for the soliton fission experiment with parameters $\alpha_3 = 0.5$, $\alpha_2 = -0.9$,   $\alpha_1 = 0.2$, $k_1 = 0.9$,     $k_2 = -0.9$, $k_3 = 1.6$, { $\phi_0 = 1.0$}, $\xi_1=\xi_2=\xi_3=0$. The gPINN, cl-PINN, and cl-gPINN algorithms provide a significant improvement over regular PINNs, reducing the equation loss by an order of magnitude in all three cases. The alpha error $\epsilon_\alpha$ is also decreased. {These values are summarized in table \ref{tab_2}. Again, this improvement is evident only when the PINN method fails to minimize effectively the alpha error. Such failure can occur by altering the initial point in the parametric space or by increasing the static learning rate.}

The enhancements provided by the gPINN, cl-PINN, and cl-gPINN approaches come at a cost, as the training time increases significantly for the same number of training epochs. To quantify this slowdown in training, we define the training time ratio or slowdown factor $tau$ as the ratio of the training time of each one of gPINN, cl-PINN and cl-gPINN over the training time with regular PINNs:
\begin{eqnarray}
    \tau_q = \frac{T_q}{T}\,, 
\end{eqnarray}
where $T$ is the total PINN training time while $T_q$ with $q=g,cl,clg$, are the training times for the gPINNs, cl-PINN and cl-gPINNs, respectively. We also introduce the relative loss improvement factor $\lambda$ defined by
\begin{eqnarray}
    \lambda_q = \frac{L_{e,min}}{L^{(q)}_{e,min}}\,, \label{rli}
\end{eqnarray}
and the relative inference improvement factor $\kappa$ defined by 
\begin{eqnarray}
    \kappa_q = \frac{\epsilon_{\alpha}}{\epsilon^{(q)}_{\alpha}}\,. \label{rai}
\end{eqnarray}
Here, $L_{e,min}$ and $\epsilon_{\alpha}$ represent the minimum equation loss and the final alpha error values obtained using the baseline PINN algorithm, respectively, while $L_{e,min}^{(q)}$ and $\epsilon_{\alpha}^{(q)}$ denote the corresponding values achieved using the enhanced methods.

In table \ref{tab_2} we provide comparisons of the four different algorithms in terms of the parameters $\tau$, $\lambda$ and $\kappa$, for the soliton fusion and fission cases, respectively. The results in table \ref{tab_2}  indicate a significant improvement in minimizing the equation loss and inferring the PDE parameters with the enhanced algorithms; however, the computational time is also increased. The increased computational time of the enhanced algorithms is due to the additional operations needed at each epoch for each one of them. The gPINN algorithm requires the computation of 5th order derivatives and mixed $t$ and $x$ derivatives, while the cl-PINN algorithm requires the numerical integration of the $\phi$ $\forall t$ as described by Eq. \eqref{Lc}. { It should be emphasized again that this enhancement is not consistent across all experiments and strongly depends on the initialization of network parameters and the learning rate. Through extensive experimentation, we confirm that improvements occur  when the standard PINN algorithm struggles to minimize both the equation loss and the alpha error. Moreover, the enhancement is highly sensitive to training hyperparameters and loss term weights.

Overall, the gPINN approach emerges as the most efficient method, demonstrating consistent relative loss improvement in soliton fusion and fission experiments, as well as superior parameter inference,  while maintaining a slowdown factor of $<2$. Although cl-PINNs and cl-gPINNs achieve similar improvements, their slowdown factors reach $\sim 10$, making gPINNs the more efficient choice. The ability of the enhanced algorithms to achieve a smaller alpha loss ($\epsilon_\alpha$) in cases where the standard PINN method struggles, suggests that they alleviate the need for extensive fine-tuning of parameters such as grid resolution, neuron count, and learning rate. This is particularly advantageous in complex scenarios where manual tuning is time-consuming. However, further testing is required to evaluate the robustness of these improvements across different soliton structures and training settings. }

\begin{table}[h!]
{
    \centering
  \begin{tabular}{llllll} 
    \toprule
    {Fusion experiment}\\
    \midrule
    {Algorithm}& $L_e$& $\epsilon_\alpha$ &{$\lambda $}& {$\kappa$}& {$\tau$}   \\
    \midrule
PINN  & $2.1\times 10^{-3}$ & $1.36$ & 1 & 1 &   1  \\
gPINN&  $3.1\times 10^{-5}$ & $0.49$ & 67.7  & 2.77 & 1.62  \\
cl-PINN& $5.5\times 10^{-4}$& $0.45$ & 3.81  & 3.02 & 9.40   \\
cl- gPINN& $5.8\times 10^{-4}$& 0.58 & 3.62 & 2.34 & 10.33   \\
    \bottomrule
\end{tabular}

  \begin{tabular}{llllll} 
    \toprule
    {Fission experiment}\\
    \midrule
    {Algorithm}&$L_e$& $\epsilon_\alpha$ &{$\lambda $}& {$\kappa$}& {$\tau$}   \\
    \midrule
PINN & $1.6\times 10^{-3}$& 2.00 & 1 & 1 &   1  \\
gPINN& $1.4\times 10^{-4}$ & 0.32 & 11.43  & 6.25 & 1.73  \\
cl-PINN& $4.9\times 10^{-4}$& 0.55 & 3.26  & 3.64 & 9.02   \\
cl- gPINN& $4.0\times 10^{-4}$& 0.92& 4.07 & 2.17 & 9.97  \\
    \bottomrule
\end{tabular}
    \caption{{The training metrics $L_e$ and $\epsilon_\alpha$, the accuracy increase measures $\lambda$ and $\kappa$ and the slowdown factor $\tau$ for the the PINN, gPINN, cl-PINN and cl-gPINN algorithms in the soliton fusion and fission cases (See Fig. \ref{fig_hist_enh_fus-fis}). The displayed values of $L_e$ corresponds to the best training epoch, while the values of $\epsilon_\alpha$ are taken from the last training epoch.}}
    \label{tab_2}}
\end{table}

\begin{figure}[ht!]
    \centering
    \includegraphics[scale=0.21]{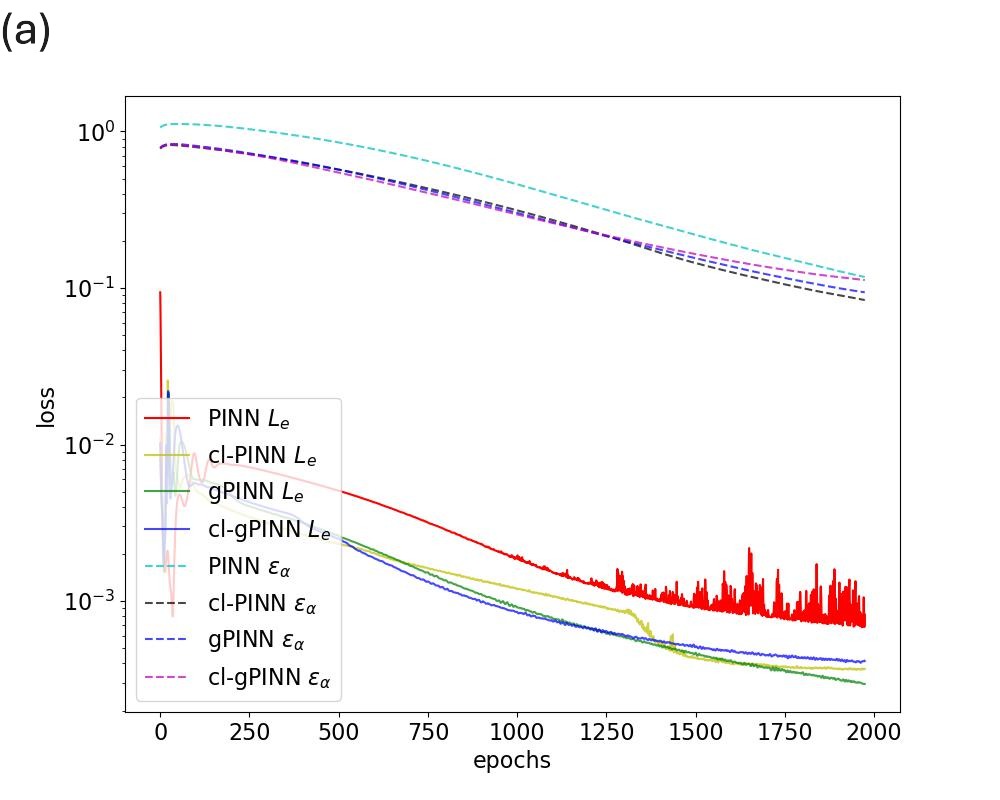}
    \includegraphics[scale=0.21]{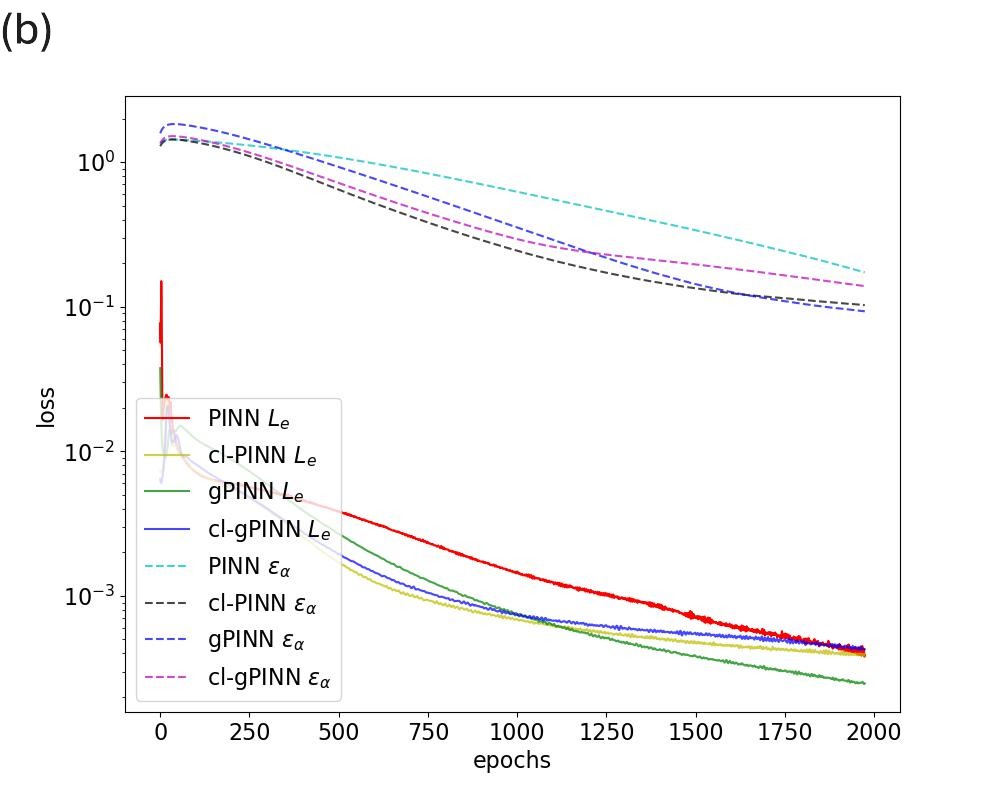}
    \caption{{Training history for the soliton  fusion (a) and soliton fission (b) numerical experiments with only one trainable parameter ($\alpha_1$). In this case the baseline PINN algorithm can infer more effectively the unknown parameter than in the previous case with three trainable parameters (Fig. \ref{fig_hist_enh_fus-fis}) and the enhanced alorithms can further improve the accuracy  by minimizing more effectively the equation loss and the alpha error.}}
    \label{fig_enhancement_simpler_case}
\end{figure}

 As a final note, we emphasize that the proposed enhancement is less critical for parametric inference in simpler inverse problems involving fewer unknown parameters and less complex PDEs, such as those studied in \cite{Raissi2019}. {To illustrate this, Figure \ref{fig_enhancement_simpler_case} shows the training history of the baseline PINN and the enhanced algorithms under the same training settings, parametric values, and hyperparameters as those used in the experiment corresponding to Fig. \ref{fig_hist_enh_fus-fis}, but with only the parameter $\alpha_1$ inferred while $\alpha_2$ and $\alpha_3$ are fixed. Clearly, the baseline PINN method achieves an order-of-magnitude smaller alpha error ($\epsilon_\alpha$) in this simpler case (See Fig. \ref{fig_enhancement_simpler_case} and Table \ref{tab_3}) compared to the more complex case where all three parameters $\alpha_1$, $\alpha_2$, and $\alpha_3$ are unknown (see Fig. \ref{fig_hist_enh_fus-fis} and Table \ref{tab_2}).

The key observation here is that while the enhanced algorithms have the tendency to provide an improvement across both scenarios, their relative benefit is more crucial in complex cases with more unknowns. In simpler cases, the baseline PINN performs reasonably well, and the additional gains from the enhanced algorithms, are less critical for parametric inference.  

}

\begin{table}[h!]
{
    \centering
  \begin{tabular}{llllll} 
  \toprule
  {Fusion experiment}\\
    \midrule
    {Algorithm}&$L_e$& $\epsilon_\alpha$ &{$\lambda $}& {$\kappa$}& {$\tau$}   \\
    \midrule
PINN & $7.1\times 10^{-4}$& 0.12 & 1 & 1 &   1  \\
gPINN& $2.9\times 10^{-4}$ & 0.09 &  2.45 & 1.33 & 2.11  \\
cl-PINN& $3.9\times 10^{-4}$& 0.10 & 1.82  & 1.20 & 10.72   \\
cl- gPINN& $4.0\times 10^{-4}$& 0.11& 1.78 & 1.09 & 12.34  \\
    \bottomrule
\end{tabular}

  \begin{tabular}{llllll} 
    \toprule
    {Fission experiment}\\
    \midrule
    {Algorithm}&$L_e$& $\epsilon_\alpha$ &{$\lambda $}& {$\kappa$}& {$\tau$}   \\
    \midrule
PINN & $4.1\times 10^{-4}$& 0.17 & 1 & 1 &   1  \\
gPINN& $2.5\times 10^{-4}$ & 0.09 & 1.64  & 1.89 & 2.04  \\
cl-PINN& $3.9\times 10^{-4}$& 0.10 & 1.05  & 1.70 & 11.03   \\
cl- gPINN& $4.3\times 10^{-4}$& 0.13& 0.95 & 1.31 & 11.41  \\
    \bottomrule
\end{tabular}
    \caption{{The training metrics $L_e$ and $\epsilon_\alpha$, the accuracy increase measures $\lambda$ and $\kappa$ and the slowdown factor $\tau$ for the the PINN, gPINN, cl-PINN and cl-gPINN algorithms in the soliton fusion and  fission cases with only one learnable PDE parameter (viz. $\alpha_1$) (See Fig. \ref{fig_enhancement_simpler_case}). The value of $L_e$ corresponds to the best training epoch, while the value of $\epsilon_\alpha$ is taken from the last epoch.}}
    \label{tab_3}}
\end{table}

\section{Conclusions}
\label{sec_V}

In this study, we considered a hierarchy of nonlinear PDEs that consists of linear combinations of equations within the integrable Burgers' hierarchy. We introduced new classes of three-soliton analytic solutions to the fourth-order member of this generalized hierarchy and used Physics Informed Neural Networks to approximate these soliton solutions. We adopted a semi-supervised approach that integrates the
PDE with relevant data, effectively determining its parameters, thus
identifying the specific equation within the generic hierarchy that
best describes the data. Our findings show that PINNs can accurately model multi-soliton solutions, demonstrating their potential for solving higher-order nonlinear PDEs of the Burgers hierarchy. We also observed
the robustness of the DNN solutions for increased data noise, although
the parameter inference accuracy deteriorated at higher noise levels.
Additionally, we provided evidence of the predictive capabilities of the
DNN solutions at future times. Furthermore, we enhanced PINNs with
gradient information and a conservation law specific to the generic
Burgers’ hierarchy. These modifications can improve training and inference accuracy only in certain cases. In terms of efficiency only gPINNs provided efficiency gains, as the computational time for cl-PINNs and cl-gPINNs increased significantly without offering additional accuracy improvements over the gPINN approach. { To further validate these results, additional investigation is needed. A potential direction is to compare the gPINN method with the conservation-law-constrained approach, as explored in previous works \cite{Fang2022, Wu2022}, which demonstrated
significant improvements. We leave this for future work. }

\section*{Data availability}
The data and code will be made available upon request.


\begin{appendices}

\section{Additional classes of triple-soliton solutions}
\label{appendix_A}

Classes $u_{3}^{(m)}$ with $m=2,...,6$  are given by \eqref{triple_soliton_m} each having different values of the coefficients $c$ and different dispersion relations. The different cases are listed below:
\begin{itemize}
    \item $u_3^{(2)}$: $c_{12} = c_{13} =c_{23} = c_{123} = 0$ and
\begin{eqnarray}
    \omega_2 &=& \alpha_1 k_1^2 + \alpha_2 k_1^3 + \alpha_3 k_1^4 - \alpha_1 k_2^2 - \alpha_2 k_2^3 - \alpha_3 k_2^4 + 
  \omega_1\,, \nn \\
  \omega_3 &=& \alpha_1 k_1^2 + \alpha-2 k_1^3 + \alpha_3 k_1^4 - \alpha_1 k_3^2 - \alpha_2 k_3^3 - \alpha_3 k_3^4 + 
  \omega_1\,.\nn
\end{eqnarray}
    \item $u_3^{(3)}$: $c_{12} = c_{23} =  c_{123} =0$  and
\begin{eqnarray}
\omega_1 &=& -k_1 (\alpha_1 (k_1 + 2 k_3) + \alpha_3 (k_1 + 2 k_3) (k_1^2 + 2 k_1 k_3 + 2 k_3^2) +
      \alpha_2 (k_1^2 + 3 k_1 k_3 + 3 k_3^2)),  \nn \\ 
 \omega_2 &=& -\alpha_1 (k_2^2 + 2 k_1 k_3) - \alpha_2 (k_2^3 + 3 k_1 k_3 (k_1 + k_3)) - 
   \alpha_3 (k_2^4 + 2 k_1 k_3 (2 k_1^2 + 3 k_1 k_3 + 2 k_3^2)), \nn \\
 \omega_3 &=& -k_3 (\alpha_1 (2 k_1 + k_3) + 
     \alpha_3 (2 k_1 + k_3) (2 k_1^2 + 2 k_1 k_3 + k_3^2) + 
     \alpha_2 (3 k_1^2 + 3 k_1 k_3 + k_3^2))\,. \nn 
\end{eqnarray}
    \item $u_3^{(4)}$:      $c_{12} = c_{13} = c_{123} = 0$  and
\begin{eqnarray}
\omega_1 &=& -\alpha_1 (k_1^2 + 2 k_2 k_3) - \alpha_2 (k_1^3 + 3 k_2 k_3 (k_2 + k_3)) - 
  \alpha_3 (k_1^4 + 
     2 k_2 k_3 (2 k_2^2 + 3 k_2 k_3 + 2 k_3^2)), \nn\\
     \omega_2 &=& -k_2 (\alpha_1 (k_2 + 2 k_3) +
     \alpha_3 (k_2 + 2 k_3) (k_2^2 + 2 k_2 k_3 + 2 k_3^2) + 
    \alpha_2 (k_2^2 + 3 k_2 k_3 + 3 k_3^2))\nn\\
    \omega_3 &=& -k_3 (\alpha_1 (2 k_2 + k_3) + 
    \alpha_3 (2 k_2 + k_3) (2 k_2^2 + 2 k_2 k_3 + k_3^2) + 
    \alpha_2 (3 k_2^2 + 3 k_2 k_3 + 
       k_3^2))\,. \nn 
\end{eqnarray}
    \item $u_3^{(5)}$: $ c_{13} =  c_{23} = c_{123} = 0$ and
\begin{eqnarray}
\omega_1 &=& -k_1 (\alpha_1 (k_1 + 2 k_2) + 
     \alpha_3 (k_1 + 2 k_2) (k_1^2 + 2 k_1 k_2 + 2 k_2^2) + 
     \alpha_2 (k_1^2 + 3 k_1 k_2 + 3 k_2^2)), \nn \\
\omega_2 &=& -k_2 (\alpha_1 (2 k_1 + k_2) + 
     \alpha_3 (2 k_1 + k_2) (2 k_1^2 + 2 k_1 k_2 + k_2^2) + 
     \alpha_2 (3 k_1^2 + 3 k_1 k_2 + k_2^2)), \nn \\ 
\omega_3 &=& -\alpha_1 (2 k_1 k_2 + k_3^2) - \alpha_2 (3 k_1 k_2 (k_1 + k_2) + k_3^3) - 
   \alpha_3 (2 k_1 k_2 (2 k_1^2 + 3 k_1 k_2 + 2 k_2^2) + k_3^4)\,.\nn
\end{eqnarray}

    \item $u_3^{(6)}$: $  c_{12} =  c_{13} = c_{23} = 0$ and
\begin{eqnarray}
\omega_1 &=& -\alpha_1 (k_1 + k_2) (k_1 + k_3) - 
  \alpha_3 (k_1 + k_2) (k_1 + k_3) (k_1^2 + 2 k_2^2 + 3 k_2 k_3 + 2 k_3^2  \nn \\ 
    &&+ 
     k_1 (k_2 + k_3)) - 
  \frac{1}{2} \alpha_2 (2 k_1^3 + 3 k_1^2 (k_2 + k_3) + 3 k_2 k_3 (k_2 + k_3) + 
     3 k_1 (k_2 + k_3)^2), \nn \\   
     \omega_2 &=& -\alpha_1 (k_1 + k_2) (k_2 + k_3) - 
  \alpha_3 (k_1 + k_2) (k_2 + k_3) (2 k_1^2 + k_2^2 + k_2 k_3 + 2 k_3^2  \nn \\ 
  &&+ 
     k_1 (k_2 + 3 k_3))- \frac{1}{2} \alpha_2 (3 k_1^2 (k_2 + k_3) + 3 k_1 (k_2 + k_3)^2 + 
     k_2 (2 k_2^2 + 3 k_2 k_3 + 3 k_3^2)), \nn \\
     \omega_3 &=& -\alpha_1 (k_1 + k_3) (k_2 + k_3) - 
  \alpha_3 (k_1 + k_3) (k_2 + k_3) (2 k_1^2 + 3 k_1 k_2 + 2 k_2^2 + (k_1 + k_2) k_3  \nn \\ 
     &&+ 
     k_3^2)+ \frac{1}{2} \alpha_2 (-3 k_1 k_2 (k_1 + k_2) - 3 (k_1 + k_2)^2 k_3 - 3 (k_1 + k_2) k_3^2 - 
     2 k_3^3)\,.\nn
\end{eqnarray}
\end{itemize}
Notice that the classes $u_3^{(3)}$, $u_{3}^{(4)}$ and $u_{3}^{(5)}$ are essentially equivalent as each class can be recovered by the others by permuting the indices $3,4,5$. 
\end{appendices}

\printbibliography

\end{document}